\documentclass[preprint,12pt,authoryear]{elsarticle}
\usepackage{epsfig}
\usepackage{graphicx}
\usepackage{amsmath}
\usepackage{amssymb}
\usepackage{amsfonts}
\usepackage[geometry]{ifsym}
\usepackage{url}
\usepackage{siunitx}
\usepackage{graphicx}
\usepackage{blindtext}
\usepackage{scrextend}
\usepackage{subfig}
\usepackage{algorithmic} 
\usepackage{threeparttable}
\usepackage{verbatim}
\usepackage{epstopdf}
\usepackage{todonotes}
\usepackage{hyperref}
\usepackage{natbib}

\newcommand{\norme}[1]{\left\Vert #1\right\Vert}
\newcommand{\vv}[1]{\mathbf{#1}}

\makeatletter
\def\blfootnote{\xdef\@thefnmark{}\@footnotetext}
\makeatother
\addtokomafont{labelinglabel}{\sffamily}

\journal{IEEE Transactions on CST}

\begin{document}

\begin{frontmatter}
%
\title{$\mathcal{L}_2$ and $\mathcal{L}_\infty$ stability analysis of heterogeneous traffic with application to parameter optimisation for the control of automated vehicles}
%
%
%

\title{$\mathcal{L}_{2}$ and $\mathcal{L}_{\infty}$ stability analysis of heterogeneous traffic with application to parameter optimisation for the control of automated vehicles}

\author{Julien Monteil\fnref{label}}
\author{M\'elanie Bouroche\fnref{labelTri}}
\author{Douglas J. Leith\fnref{labelTri}}
\fntext[label]{IBM Research - Ireland Lab, Control and Optimization Group. Address: Building $3$, IBM Technology Campus, Mulhuddart, Dublin~$15$, Ireland.}
\fntext[labelTri]{Trinity College Dublin - School of Computer Science. Address: College Green, Dublin~$2$, Ireland.}


\begin{abstract} The presence of (partially) automated vehicles on the roads presents an opportunity to compensate the unstable behaviour of conventional vehicles. Vehicles subject to perturbations should (i) recover their equilibrium speed, (ii) react not to propagate but absorb perturbations. In this work, we start with considering vehicle systems consisting of heterogeneous vehicles updating their dynamics according to realistic behavioural car-following models. Definitions of all types of stability that are of interest in the vehicle system, namely input-output stability, scalability, weak and strict string stability, are introduced based on recent studies. Then, frequency domain linear stability analyses are conducted after linearisation of the modelled system of vehicles, leading to conditions for input-output stability, strict and weak string stability over the behavioural parameters of the system, for finite and infinite systems of homogeneous and heterogeneous vehicles. This provides a solid basis that was missing for car-following model-based control design in mixed traffic systems where only a proportion of vehicles can be controlled. After visualisation of the theoretical results in simulation, we formulate an optimisation strategy with LMI constraints to tune the behavioural parameters of the automated vehicles in order to maximise the $\mathcal{L}_\infty$ string stability of the mixed traffic flow while considering the comfort of automated driving. The optimisation strategy systematically leads to increased traffic flow stability. We show that very few automated vehicles are required to prevent the propagation of realistic disturbances. 
\end{abstract}


\end{frontmatter}

%

\section{Introduction}

Interest is growing in how to control (partially) automated vehicles, i.e. vehicles equipped with Automated Driving Systems (ADS) such as Adaptive Cruise Control (ACC), Cooperative Adaptive Cruise Control (CACC), or any autopilot system, to increase traffic flow stability and safety in mixed traffic contexts, when (partially) automated vehicles and conventional vehicles coexist on the road. A recent report has underlined the unsafe nature of automated vehicles, which are five times more likely to crash than conventional vehicles, even though they are almost never to blame when a crash does occur~\cite{UMTRI}. In this context, one can assume that automated vehicles should behave similarly to surrounding drivers in order not to surprise them and, when automation is only partial, similarly to the in-vehicle driver in order to increase driving comfort and facilitate switching between automated and conventional modes. As a result, the modelling and understanding of conventional vehicles dynamics is particularly relevant for the design of driver-dependent, comfortable, and safe controllers for the acceleration dynamics of automated vehicles. 

Regarding automated vehicle dynamics, a lot of attention has been put into the study of platoon systems, i.e. systems composed of automated vehicles that behave according to some distributed control protocol. Two stability objectives, namely stability and string stability, are commonly considered when designing distributed control protocols for some given communication topologies, see e.g.~\cite{knorn2014passivity,6515636}. In particular, while stability characterises the convergence of the platooning system towards a desired equilibrium configuration, string stability refers to the attenuation of disturbances along the vehicle string. Note that, despite a number of early works~\cite{4790743,260756,ZHOUJing:30,5571043}, the lack of a unified definition for string stability was identified in~\cite{6515636}, where a definition of $\mathcal{L}_p$ string stability was proposed for interconnected systems, and several of its implications were discussed. In the field of platoon systems, pioneering work was made in the context of the Automated Highway System (AHS) effort of the California Path Program~\cite{shladover1991automated,varaiya1991sketch}. Among the contributions were the formulations of lateral and longitudinal control laws for stable, safe and comfortable platoon maneuvres, see~\cite{li1997ahs,frankel1996safety} for example, the development of a control system architecture for platooning, see~\cite{hedrick1994control} for instance, and the San Diego demonstration in 1997~\cite{tan1998demonstration}. Since then, several linear control protocols have been proposed over the years to address the stability and string stability of platoons, see e.g.~\cite{swaroop1995string,liu2001effects}. Very recently, the string stable behaviour of a linear CACC system via $\mathcal{H}_{\infty}$ control was demonstrated~\cite{ploeg2014controller}, the stable platooning of a linear control protocol considering communication delays was proved using the Lyapunov-Razumikhin theorem~\cite{6891349}, a novel condition for the design of nonlinear control protocols for stable platooning was proposed~\cite{7937859}, and a protocol with nonlinear control terms was proved to lead to string stable platooning via energy-based arguments~\cite{knorn2014passivity}. 

Regarding conventional vehicle dynamics, the longitudinal microscopic behaviour of vehicles is usually approximated using car-following models, where vehicles only react to the behaviour of their leaders. Car-following models are continuous time models that can be either delay-free~\cite{kesting1,treiber1} or delayed~\cite{newell,rakha}. These models have been shown to accurately represent traffic flow features such as speed and headway distributions as well as stop and go waves, using real world datasets~\cite{treiber2,monteil1,punzo1}, and are therefore a solid base for further developments on stability analysis and controller synthesis. The string stability characterisation of these models has received some recent attention. One key work is the one of~\cite{wilson}, where the behaviour of a uniform flow of vehicles with car-following dynamics was investigated. After linearisation of the dynamics of the infinite homogeneous system it is shown via Fourier analysis that conventional vehicles can have a string stable or string unstable behaviour depending on the traffic regime and their behavioural parameters. One of the early works to use a similar approach appeared in~\cite{Bando}, and this was recently utilised to study weakly non-linear congestion patterns of bilateral multi-anticipative traffic~\cite{monteil2} as well as multi-anticipative traffic introducing time-delays~\cite{Ngoduy2015420}. However, the string stability of car-following dynamics remains not well understood in the case of finite systems of homogeneous vehicles as well as in the case of systems of heterogeneous vehicles, i.e. with vehicles exhibiting different car-following behaviour. 

\subsubsection*{Contributions of the paper} in the context of the above literature, we propose a number of key novelties and contributions. First, we believe that the key and novel idea of this paper is that in mixed traffic environments (partially) automated vehicles can be used to compensate the string unstable behaviour of conventional vehicles. Differently to the problem of designing control protocols for automated platoon systems, here we focus on mixed vehicle systems where the dynamics of conventional vehicles are heterogeneous, follow realistic car-following behaviour and cannot be controlled. Furthermore, we work with the hypothesis that the automated vehicles should behave in a similar way to the conventional vehicles to maximise driving comfort and compliance with the automated system, and to minimise the occurrences of unusual behaviour. In such a context, the contributions in comparison with existing works are as follows: (i) we formalise the definition of weak stability that is relevant in mixed traffic environments made of automated and conventional vehicles; (ii) we investigate the stability of the heterogeneous systems of vehicles with linearised car-following dynamics in the frequency domain, making use of $\mathcal{L}_2$ linear control theory, which is widely used in the platoon literature; (iii) we provide conditions for input-output and string stability of heterogeneous traffic, for single and multiple considered outputs; (iv) we provide a relation between $\mathcal{L}_2$ and $\mathcal{L}_{\infty}$ string stability for the considered dynamics; (v) we show that the equivalence between string stability and asymptotic stability does not hold for closed loop systems; (vi) we extensively discuss those results in simulation, showing the critical features of nonlinearities; (vii) based on the weak stability condition, we propose an optimisation strategy to tune the behavioural parameters of the automated vehicles, which has a Linear Matrix Inequalities (LMI) formulation; (viii) we apply the optimisation strategy to realistic data which yields very promising results: a very small proportion of automated vehicles can greatly and systematically contribute to increasing traffic flow stability. 

\subsubsection*{Organisation of the paper} Section~\ref{1} lists the notation used in the paper. Section~\ref{2} recalls the general form of car-following models, and describes the linearisation of the heterogeneous vehicle system. Section~\ref{3} presents the definition of input-output stability, strict and weak string stability for heterogeneous systems of vehicles. We introduce the definition of weak stability, relevant to mixed traffic environments. In Section~\ref{4}, the definitions are applied to the system of vehicles providing a list of necessary and sufficient conditions when possible, sufficient conditions otherwise, for input-output, strict and weak string stability, for both homogeneous and heterogeneous traffic. In Section~\ref{5}, we visualise these results in simulation, and discuss linear vs non-linear stability. In Section~\ref{6}, we provide an optimisation strategy to tune the behavioural parameters of the (partially) automated vehicles in the traffic system so as to increase weak string stability while considering the comfort of driving. We formulate the constraints as LMI, and solve the optimisation using its convex structure. We show that using our approach automated vehicles consistently contribute to increasing traffic flow stability of heterogeneous traffic. We conclude with a summary of our findings. 

\section{Notations}\label{1}

The notation used throughout this paper is as follows. 

\begin{labeling}{alligator}
\item [$\mathbb{N}$] Set of natural numbers.
\item [$\mathbb{R}$] Set of real numbers.
\item [$\mathbb{R}_+$] Set of positive real numbers.
\item [$\mathbb{R}_+^*$] Set of strictly positive real numbers.
\item [$\mathbb{R}^{p\times q}$] Set of matrices with coefficients in $\mathbb{R}$, $p$ and $q\in \mathbb{N}$.
\item [$||.||_{2}$] $L_2$ norm of a vector in $\mathbb{R}^n$, with $n\in \mathbb{N}$. 
\item [$\mathcal{L}_p$] Space of functions $f:\mathbb{R}\rightarrow\mathbb{R}$ such that $t\rightarrow|f(t)|^p$ is integrable over $\mathbb{R}$, here $p=2,\infty$.
\item [$||.||_{\mathcal{L}_p}$] $\mathcal{L}_p$ norm of a $\mathcal{L}_{p}$ function, here $p=2,\infty$.
\item [$Y_i(s)$] Laplace transform of $y_i(t)$.
\item [$D_i(s)$] Laplace transform of $d_i(t)$.
\item [$||.||_{\mathcal{\mathcal{H}}_{\infty}}$] $\mathcal{H}_{\infty}$ norm of a defined Laplace transform.
\item [$Re(z)$] Real part of complex number $z$.
\item [$\mathcal{K}$] Class of continuous functions $h$, defined such as $h(\cdot): \mathbb{R}_+ \rightarrow \mathbb{R}_+$, $h(0) = 0$ and $h(\cdot)$ is strictly increasing.
\end{labeling}

\section{General form of car-following models and corresponding system equation}\label{2}

Consider a system of $m>1$ vehicles with indices $n\in\{1,...,m\}$. The first vehicle of the vehicle system ($n=1$) follows a virtual reference vehicle ($n=0$) which keeps a constant speed $\dot x_{0}(t)=v_{eq}$. The other vehicles ($1\leq n\leq m$) behave according to a car-following model. 

\subsection{Car-following models}
Car-following models describe the dynamics of a vehicle in response to the trajectory of its leading vehicle, taking into account the technical features of the car and the behaviour of the driver through vehicle behavioural parameters. 
\subsubsection{Selection of the state vector} 

before presenting the models we introduce $x_n\in \mathbb{R}^2$ the state vector of vehicle $n$ defined as:
\begin{equation}
\vv{x}_n=\begin{pmatrix}\Delta x_n\\v_n\end{pmatrix},\label{eq:sty0}
\end{equation}
where $\Delta x_n=x_{n-1}-x_n$ is the space headway or distance between vehicle $n$ and its leading vehicle $n-1$, and $v_n$ is the speed of vehicle $n$.

\subsubsection{Delay-free car-following models} delay-free continuous time models are the most well-known type of car-following models in the literature. For a vehicle $n$, the vehicle dynamics are as follows:
\begin{align}
\Delta \dot x_n&=v_{n-1}-v_n\label{CF2},\\
\dot v_n&=f_n(v_{n},\Delta x_{n}, v_{n-1}-v_n,\theta_n)+d_n,\label{CF1}
\end{align}
where $\dot v_n$ is the acceleration of the vehicle $n\in \{1,...,m\}$, $\Delta x_{n}=x_{n-1}-x_{n}$ the distance to the vehicle in front (headway), $\Delta \dot x_{n}=\dot x_{n-1}-\dot x_{n}$ the relative velocity, \textbf{$\theta$}$_n\in \mathbb{R}^l$ the vector of the behavioural parameters, with $l\in \mathbb{N}^*$ the number of parameters, and $f_n$ the acceleration function of the car following model which captures the non linear dynamics of the vehicle. The term $d_n$ captures the effect of external disturbances. 

Note that, in this paper, we restrict our analysis to delay-free car-following models, but the presented approach could be readily extended to the consideration of delayed car-following models.  

%

\subsection{Linearisation of the vehicle state}
 
For any vehicle $n\in\{1,...,m\}$, an acceleration input perturbation on vehicle $i\in\{1,...,n\}$ generates responses $\dot{x}_{n}$, $\Delta x_{n}$, $\Delta\dot{x}_{n}$ that can be written as perturbations about the equilibrium values $\dot{x}_{n,\text{eq}}$, $\Delta x_{n,\text{eq}}$, $\Delta\dot{x}_{n,\text{eq}}$
\begin{eqnarray}
x_n(t)&=&x_{n,eq}(t)+y_n(t),\\
\dot{x}_n(t)&=&\dot{x}_{n,\text{eq}}+\dot{y}_n(t),\label{Veq}\\
\Delta x_n(t)&=&\Delta x_{n,\text{eq}}(t)+\Delta y_{n}(t),
\end{eqnarray}
where perturbations $y_n(t)=x_n(t)-x_{n,eq}(t)$ and $\Delta y_n(t)=y_{n-1}(t)-y_n(t)$. At equilibrium, we also have $\Delta \dot x_{n,\text{eq}}=0$, $\dot{x}_{n,\text{eq}}=v_{\text{eq}}$ and $\ddot{x}_n(t)=\ddot{y}_n(t)$. When $\dot{y}_n$, $\Delta y_n$, and $\Delta \dot{y}_n$ are small enough, the following approximation is valid: 
\begin{equation}
\ddot y_n \approx \dot{y}_nf_{n,1}+\Delta y_nf_{n,2}+\Delta \dot{y}_nf_{n,3}+d_n\label{L2},
\end{equation}
where $f_{n,1}=\partial f_n(\dot{x}_{n,\text{eq}},\Delta x_{n,\text{eq}},0)/\partial \dot{x}_n$, $f_{n,2}=\partial f_n(\dot{x}_{n,\text{eq}},\Delta x_{n,\text{eq}},0)/\partial \Delta x_n$, and $f_{n,3}=\partial f_n(\dot{x}_{n,\text{eq}},\Delta x_{n,\text{eq}},0)/\partial \Delta \dot{x}_n$.
Note that, as mentioned in~\cite{wilson}, equation~\eqref{L2} has a physical meaning in traffic. When the speed deviation $\dot{y}_n$ is increased, the vehicle tends to decelerate to return to the equilibrium speed; when the relative distance $\Delta {y}_n$ or relative speed $\Delta \dot{y}_n$ is increased, it tends to accelerate to return to the equilibrium speed. These considerations lead to the following conditions: $\forall n\in \{1,...,m\}$,
\begin{equation}
\ f_{n,1}<0,\ f_{n,2}>0,\ f_{n,3}>0.\label{rat}
\end{equation}

\subsection{State-space dynamics}

By analogy with~\eqref{eq:sty0}, we have $\vv{y}_n\in \mathbb{R}^2$ the perturbed state vector defined as:
\begin{equation}
\vv{y}_n=\begin{pmatrix}\Delta y_n\\\dot{y}_n\end{pmatrix}.\label{eq:sty1}
\end{equation}
The relation \eqref{L2} can then be rewritten equivalently as:
\begin{equation}
\dot{\vv{y}}_n=\vv{a}_{n,0}\vv{y}_{n-1}+\vv{a}_{n,1}\vv{y}_{n}+\vv{b_v}d_n,\label{eq:sty2}
\end{equation}
where $d_n\in \mathbb{R}$ is the external acceleration input, $\vv{b_v}\in \mathbb{R}^2$ is the following vector,
\begin{eqnarray}
\vv{b_v}&=&\begin{pmatrix}0\\1\end{pmatrix},\label{eq:b}
\end{eqnarray}
and the $\vv{a}_{n,0}$ and $\vv{a}_{n,1}\in \mathbb{R}^{2\times2}(\mathbb{R})$ are the matrices:
\begin{eqnarray}
\vv{a}_{n,0}&=&\begin{pmatrix}0&1\\0&\ f_{n,3}\end{pmatrix},\label{eq:a0}\\
\vv{a}_{n,1}&=&\begin{pmatrix}0&-1\\f_{n,2}&\ f_{n,1}-f_{n,3}\label{eq:a1}\end{pmatrix}.
\end{eqnarray}
Note that the form of vector $\vv{b_v}$ enforces a zero speed input, as the actuator is assumed to be the throttle pedal which only acts upon acceleration.

\subsection{Linearisation of the vehicle system equation}
Let $\vv{d}\in \mathbb{R}^{m}$ be the vector of disturbances:
\begin{equation}
\vv{d}=\begin{pmatrix} d_1\\\vdots\\ d_m\end{pmatrix},\label{eq:stU}\\
\end{equation}
and $\vv{y}\in \mathbb{R}^{2m}$ the state vector:
\begin{equation}
\vv{y}=\begin{pmatrix} \vv{y}_1\\\vdots\\ \vv{y}_m\end{pmatrix}.\label{eq:stY}
\end{equation}
For the leader of the system ($n=1$), there is only a fictitious vehicle ahead ($n=0$). The equation of motion of vehicle $0$ is:
\begin{equation}
\dot{\vv{y}}_0=\tilde{\vv{a}}\vv{y}_0, \label{eq:y1}
\end{equation}
with $\tilde{\vv{a}}\in \mathbb{R}^{2\times 2}$ the matrix:
\begin{equation}
\tilde{\vv{a}}=\begin{pmatrix}0&1\\0&0  \end{pmatrix}.\label{eq:ta10}
\end{equation}
The linearised dynamics of the system are:
\begin{align}
\dot{\vv{y}}&=\vv{a}\vv{y}+\vv{b}\vv{d},\label{DefLS}\\ 
\vv{h}&=\vv{c}\vv{y},\label{ObserLS}
\end{align}
where $\vv{h}\in \mathbb{R}^{2m}$ is the vector of outputs, $\vv{c}\in\mathbb{R}^{2m\times 2m}$ is the observation matrix, and the matrix $\vv{b}\in\mathbb{R}^{2m\times 2m}$ is:
\begin{equation}
\vv{b}=\begin{pmatrix}\vv{b_v}&0&\ldots&0\\
0&\vv{b_v}&\ldots&0\\
\vdots&\ddots&\ddots&0\\
0&\ldots&0&\vv{b_v}
\end{pmatrix},\label{eq:mB}
\end{equation}
where $\vv{b_v}$ is given by equation~\eqref{eq:b} and matrix $\vv{a}\in \mathbb{R}^{2m\times 2m}$ has a $2\times 2$ block form:
\begin{equation}
\vv{a}=\begin{pmatrix}\vv{a}_{1,0}&\vv{a}_{1,1}&0&\ldots&0\\
0&\ddots&\ddots&\ddots&\vdots\\
\vdots&\ddots&\ddots&\ddots&0\\
0&\ldots&0&\vv{a}_{m,0}&\vv{a}_{m,1}
\end{pmatrix}.\label{eq:matrice}
\end{equation}

The equilibrium state of the linear system is defined as $\vv{y}_{\text{eq}}$, which is also the zero vector $\in \mathbb{R}^{2m}$ (since $\vv{d}$ is zero in equilibrium), and $\vv{h}_{\text{eq}}=\vv{c}\vv{y}_{\text{eq}}$ is the equilibrium output. $\vv{h}_n\in \mathbb{R}^{2}$ is the output of vehicle $n$, and $\vv{h}_{n,\text{eq}}\in \mathbb{R}^{2}$ is the equilibrium output of vehicle $n$. Note that the form of the observation matrix $\vv{c}$ depends on the characteristics of the observer. For a centralised observer $\vv{c}$ would be the identity matrix $\in \mathbb{R}^{2m\times 2m}$. However, a decentralised observer such as a vehicle only sees a local part of the full state vector. 
 
\section{Stability definitions and remarks}\label{3}

In this section, we assume $\vv{y}_{\text{eq}}=\vv{0}$ for the sake of simplicity.

\textit{Definition 1 (Stability and exponential stability)}, see~\cite{7170953}. Consider the linear system defined in equations~\eqref{DefLS},~\eqref{ObserLS}. The vehicle system of $m>1$ vehicles is said to be stable if for a given $t_0>0$, $\forall \epsilon>0$, $\exists \ \delta=\delta(\epsilon)>0$ such that when $||\vv{y}(t_0)||_{2}<\delta$, then $\forall t\geq t_0$, $||\vv{y}||_{\mathcal{L}_p}<\epsilon$; and the system is said to be exponentially stable if for every $\delta\ge 0$ there exists $\alpha$, $\beta \in\mathbb{R_+^{*}}$ such that if $||\vv{y}(t_0)||_{2}<\delta$, then $\|\vv{y}\|_{\mathcal{L}_p} \le \alpha\|\vv{y}(t_0)\|_{2}e^{-\beta(t-t_0)} $.

Note that in the case of the linear system of equation~\eqref{DefLS},~\eqref{ObserLS}, the system is exponentially stable iff $\forall \lambda \in \{\text{Spectrum(\textbf{a})}\}$, $\text{Re}(\lambda)<0$, see e.g.~\cite{Hespanha09}. 

\textit{Definition 2 (Input-output stability)}, see e.g.~\cite{6515636},~\cite{7937859}. Consider the linear system defined in equations~\eqref{DefLS},~\eqref{ObserLS}. The vehicle system of $m>1$ vehicles is said to be input-output stable if there exists class $\mathcal{K}$ functions $\alpha:[0,a)\mapsto [0,\infty)$ and $\beta:[0,b)\mapsto [0,\infty)$, $a$ and $b\in \mathbb{R}$, such that, for any initial state $\vv{y}(t_0)\in \mathbb{R}^{2m}$ and any input $\vv{d}\in\mathcal{L}_p$, then $\forall n\in\{1,...,m\}$,
\begin{equation}
||\vv{h}_n||_{\mathcal{L}_p}\leq \alpha (\|\vv{d}\|_{\mathcal{L}_p}) +\beta (||\vv{h}(t_0)||_{2})\label{SSS1}
\end{equation}

\textit{Remark 1}. For linear systems when input-output stability holds element-wise then it holds for all inputs, i.e. when the system is input-output stable for inputs $\vv{d}\in \mathbb{R}^{m}$ such that $d_i$, $i\in \{1,...,m\}$, is the only non-zero component of $\vv{d}$ then it also holds for all inputs (by superposition). 

Note that for fully controllable and observable linear systems, exponential stability is equivalent to input-output stability, i.e. matrix $\vv{a}$ is Hurwitz, and that for non-linear systems, exponential stability implies input-output stability, see e.g.~\cite{Hespanha09}.

\textit{Remark 2.} If the input-output property holds $\forall m\in\mathbb{N}\setminus\{1\}$, i.e. the class $\mathcal{K}$ functions $\alpha$ and $\beta$ do not depend on $m$, then from the literature the vehicle system is said to be string stable, or scalable~\cite{6515636,darbha2005information}. We will use the term scalable to avoid any confusion.

We now give the definition of ${\mathcal{L}_p}$ strict string stability, adapted from~\cite{6515636}, with $p=2$ or $p=\infty$, see Section~\ref{1}. Note that, as mentioned in~\cite{5524703} for instance, the major part of the string stability analyses in the literature deal with the ${\mathcal{L}_2}$ norm which is easier to work with, as it can rewritten immediately as a condition on the $\mathcal{H}_{\infty}$ norm of the corresponding transfer function. However, the ${\mathcal{L}_\infty}$ norm has a stronger physical meaning as it deals with the peak of the deviations, and therefore ${\mathcal{L}_{\infty}}$ strict string stability can be directly related to a condition for collision avoidance.  

\textit{Definition 3 (${\mathcal{L}_p}$ strict string stability)}.
We apply the definition of strict string stability in~\cite{6515636} to the linear system defined in equations~\eqref{DefLS},~\eqref{ObserLS}. The vehicle system of $m>1$ vehicles is said to be ${\mathcal{L}_p}$ strictly string stable if it is input-output stable and if, in addition, for inputs $\vv{d}\in \mathbb{R}^{m}$ such that $d_i$, $\forall i\in \{1,...,m\}$, is the only non-zero component and $\vv{y}(t_0)=\vv{0}$, then $\forall n\in\{i+1,...,m\}$,
\begin{equation}
||\vv{h}_n||_{{\mathcal{L}_p}}\leq||\vv{h}_{n-1}||_{{\mathcal{L}_p}} \label{SSS2}.
\end{equation}

We now formalise the definition of weak string stability which was first intuitively mentioned in~\cite{5571043}.

\textit{Definition 4 (${\mathcal{L}_p}$ weak string stability)}. Consider the linear system defined in equations~\eqref{DefLS},~\eqref{ObserLS}. Let the vehicle system of $m>1$ vehicles be input-output stable and consider inputs $\vv{d}\in \mathbb{R}^{m}$ such that $d_i$, $\forall i\in \{1,...,m\}$, is the only non-zero component and $\vv{y}(t_0)=\vv{0}$. Then, for given $l\in \{i,...,m\}$ and $n\in\{l,...,m\}$, the vehicle system is said to be $(l,n)$ weakly string stable if
\begin{equation}
||\vv{h}_n||_{{\mathcal{L}_p}}\leq||\vv{h}_{l}||_{{\mathcal{L}_p}}\label{SSS6}.
\end{equation} 

\textit{Remark 3}. For linear systems when string stability holds element-wise then it holds for all inputs, i.e. when the system is string stable for inputs $\vv{d}\in \mathbb{R}^{m}$ such that $d_i$, $i\in \{1,...,m\}$, is the only non-zero component of $\vv{d}$ then it also holds for all inputs (by superposition). 

\textit{Remark 4}. The weak string stability definition is introduced to handle mixed traffic situations where vehicle systems are composed of conventional vehicles and automated vehicles. The idea is to achieve the weak string stability condition by only acting upon the controllable automated vehicles. Note that for a given $i\in \{1,...,m-2\}$, we may have $(i,i+2)$ weak string stability, i.e. $||\vv{h}_{i+2}||_{{\mathcal{L}_2}}\leq||\vv{h}_{i}||_{{\mathcal{L}_2}}$, but strict string stability does not hold. 


\section{Stability results}\label{4}
In this section we investigate the stability properties of the linearised system dynamics. We consider an input $\vv{d}$ such that only element $d_i$ is non-zero, i.e. $d_n(t)=0$ for $n\ne i$, and note that $\vv{y}_j(t)=0$ for $j<i$ since there is no input to these vehicles and the initial conditions are zero. 

\subsection{Laplace transforms of the vehicle system dynamics}\label{sstring}

Taking Laplace transforms and rearranging, equation~\eqref{eq:sty2} yields: 
\begin{equation}
\vv{Y}_i(s)=(s\vv{I}-\vv{a}_{i,1})^{-1}\vv{b_v}D_i(s), 
\end{equation}
i.e.
\begin{equation}
\vv{Y}_i(s)=\vv{G}_i(s)D_i(s)\label{inpout}, 
\end{equation}  
where $\vv{I}\in\mathbb{R}^{2\times 2}$ is the identity matrix, and where transfer function $\vv{G}_i(s)$ is
\begin{equation}
\vv{G}_i(s)=\frac{1}{U_i(s)}\begin{pmatrix}
-1\\
s
\end{pmatrix},
\label{globalinput}
\end{equation}
where
\begin{equation} 
U_i(s)=s^2+s(f_{i,3}-f_{i,1})+f_{i,2}.
\end{equation}

For vehicles $n> i$ we also have
\begin{equation}
\vv{Y}_n(s)=\vv{\Gamma}_n(s)\vv{Y}_{n-1}(s) =\left(\prod_{j=i+1}^n\vv{\Gamma}_j(s)\right)\vv{Y}_i(s),
\label{stringstab}
\end{equation}
where 
\begin{equation}
\vv{\Gamma}_n(s)=(s\vv{I}-\vv{a}_{n,1})^{-1}\vv{a}_{n,0}\label{ssgeneral}.
\end{equation} 
Hence, 
\begin{equation}
\vv{Y}_n(s)=\left(\prod_{j=i+1}^n\vv{\Gamma}_j(s)\right)\vv{G}_i(s)D_i(s).\label{inpout2}
\end{equation}

\subsection{Input-output stability and scalability}

We know that, e.g. see page 26 of~\cite{Desoer}, the $\mathcal{L}_2$-induced norm of the linear map $d_i\rightarrow \vv{y}_i$ is the $\mathcal{H}_\infty$ norm of $\vv{G}_i$. By definition, the $\mathcal{H}_\infty$ norm of $\vv{G}_{i}$ is 
\begin{equation}
||\vv{G}_{i}||_{\mathcal{H}_\infty}=\sup\limits_{\omega\in\mathbb{R}}\sigma_{\max}(\vv{G}_i(j\omega)),\label{hinf}
\end{equation} 
where $\sigma_{\max}$ is the maximum singular value of the matrix. The inequality
\begin{equation}
||\vv{y}_i||_{\mathcal{L}_2}\leq ||\vv{G}_i(s)||_{\mathcal{H}_\infty}||d_i||_{\mathcal{L}_2}\label{ine1}
\end{equation} 
holds, and $||\vv{y}_i||_{\mathcal{L}_2}$ can be made equal to $||\vv{G}_i(s)||_{\mathcal{H}_\infty}||d_i||_{\mathcal{L}_2}$ by appropriate selection of input $||\vv{d}_i||_{\mathcal{L}_2}$. In addition, for $n>i$, the inequality 
\begin{equation}
||\vv{y}_n||_{\mathcal{L}_2}\leq ||\vv{\Gamma}_n||_{\mathcal{H}_\infty}||\vv{y}_{n-1}||_{\mathcal{L}_2}\label{submu1}
\end{equation}
follows from equation~\eqref{stringstab}, and we have,
\begin{equation}
||\vv{y}_n||_{\mathcal{L}_2}\leq \prod_{j=i+1}^{n}\norme{\vv{\Gamma}_j}_{\mathcal{H}_\infty}||\vv{G}_i||_{\mathcal{H}_\infty}||d_i||_{\mathcal{L}_2}\label{36}.
\end{equation}
Therefore, the existence of $\max_{i\in\mathbb{N}^*}||\vv{G}_i(s)||_{\mathcal{H}_\infty}$ and $\max_{i\in\mathbb{N}^*}||\vv{\Gamma}_i(s)||_{\mathcal{H}_\infty}$ is a sufficient condition for input-output stability of the vehicle system. In addition, with regard to Remark 2, the existence of $\sup_{i\in\mathbb{N}^*}||\vv{G}_i(s)||_{\mathcal{H}_\infty}$ and $\sup_{i\in\mathbb{N}^*}\prod_{j=2}^{i}\norme{\vv{\Gamma}_j}_{\mathcal{H}_\infty}$ is a sufficient condition for the scalability of the vehicle system. Finally, in the case of an homogeneous system with Single Input Single Output (SISO) transfer functions, we recall that  the conditions $||G_1||_{\mathcal{H}_\infty}$ bounded and $||\Gamma_2||_{\mathcal{H}_\infty}\leq 1$ become necessary and sufficient conditions for scalability, see the work of~\cite{6515636}.

In the case of our system, following equations~(\ref{DefLS}-\ref{eq:matrice}), the system matrix $\vv{a}$ is block diagonal and so the eigenvalues of $\vv{a}$ are the eigenvalues of the $m$ block matrices on the diagonal. The eigenvalues of the matrices $(\vv{a}_{n,1})_{1\leq n\leq m}$ are $\left(f_{n,1}-f_{n,3}\pm\sqrt{\Delta_n}\right)/2$, where $\Delta_n=(f_{n,1}-f_{n,3})^2-4f_{n,2}$. If we consider the realistic driving constraints of equation~\eqref{rat} to be satisfied, the eigenvalues have negative real parts, i.e. $(\vv{a}_{n,1})_{1\leq n\leq m}$ are Hurwitz matrices, and therefore the system is always exponentially stable. 

\subsection{String stability of the vehicle system}\label{strictweak}

We start the discussion with $\mathcal{L}_2$ string stability before addressing $\mathcal{L}_{\infty}$ string stability.

\subsubsection{$\mathcal{L}_2$ strict string stability} 
\begin{figure}[h!]
\includegraphics[width=\textwidth]{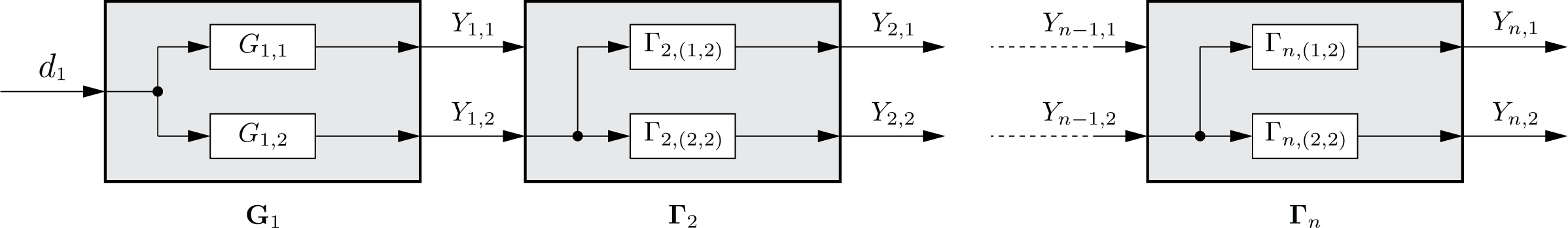}
\caption{MIMO and SISO transfer functions of the studied cascaded system.}\label{casc}
\end{figure}

applying Definition 3, if we have input-output stability, a sufficient condition for $\mathcal{L}_2$ strict string stability follows from equation~\eqref{submu1}, as mentioned in~\cite{260756}:
\begin{equation}
||\vv{\Gamma}_n||_{\mathcal{H}_\infty}\leq 1, \ \forall n\in\{2,...,m\}.\label{submu2}
\end{equation}
In the case of our system, following equations~(\ref{DefLS}-\ref{eq:matrice}), we obtain by developing equation~\eqref{ssgeneral}:
\begin{equation}
\vv{\Gamma}_n(s)=\frac{1}{U_n(s)}\begin{pmatrix}
0&s-f_{n,1}\\
0&sf_{n,3}+f_{n,2}
\end{pmatrix}.\label{weakstr}
\end{equation}
The structure of the cascaded system can be summarised in Figure~\ref{casc} for a perturbation $d_1$ on the first vehicle. $G_{1,1}$ and $G_{1,2}$ are the elements of $\vv{G}_{1}$, see equation~\eqref{globalinput}. $\vv{Y}_n$ is the Laplace transform of $\vv{y}_n$. We write $Y_{n,1}$ and $Y_{n,2}$ the Laplace transforms of $\Delta y_n$ and $\dot y_n$, and $\Gamma_{n,1}$ and $\Gamma_{n,2}$ the SISO headway to headway and speed to speed transfer functions. We have:
\begin{equation}
Y_{n,1}=\Gamma_{n,1}Y_{n-1,1}, \quad Y_{n,2}=\Gamma_{n,2}Y_{n-1,2},\label{ynk1}
\end{equation} 
Finally, $\Gamma_{n,(1,2)}$ and $\Gamma_{n,(2,2)}$ are the terms of the second column in equation~\eqref{weakstr}, and we have:
\begin{equation}
Y_{n,1}=\Gamma_{n,(1,2)}Y_{n-1,2}, \quad Y_{n,2}=\Gamma_{n,(2,2)}Y_{n-1,2}.\label{ynk2}
\end{equation}

In the rest of the Section, we are interested in capturing the propagation of the headway and speed perturbations, which can be done by looking at either the MIMO transfer function or the SISO transfer functions. 

\textit{a) Multiple Inputs Multiple Outputs (MIMO) system:} the full MIMO system is represented by transfer function $\vv{\Gamma}_n$, see Figure~\ref{casc}. We calculate the singular values of $\vv{\Gamma}_n(j\omega)$ to obtain its $\mathcal{H}_\infty$ norm, following the definition of equation~\eqref{hinf}.
After some manipulation, see Appendix A, sufficient conditions for strict string stability are: 
\begin{align}
f_{n,1}&=0\label{cond2},\\
-2f_{n,2}-1&\geq 0\label{cond3}.
\end{align}
It can be seen that the sufficient conditions for strict string stability of the MIMO system are not useful in practice, as they are not compatible with the realistic driving constraints presented in equation~\eqref{rat}.

\textit{b) Single Input Single Output (SISO) systems:} given the conservativeness of equations~\eqref{cond2},~\eqref{cond3}, the strict string stability analysis of the SISO systems is critical. From equation~(\ref{weakstr}-\ref{ynk2}) we immediately have the speed to speed transfer function:
\begin{equation}
\Gamma_{n,2}=\frac{sf_{n3}+f_{n2}}{s^2+s(f_{n,3}-f_{n,1})+f_{n,2}}.\label{tfunc01}
\end{equation}
However, the propagation of the headway perturbations is not readily obtainable, and following equation~\eqref{ynk2}, we need to express $Y_{n-1,1}$ as a function of $Y_{n-1,2}$ to have an expression of $\Gamma_{n,1}$. As we remark that $Y_{n-1,2}-Y_{n,2}=sY_{n,1}$, following equation~\eqref{ynk1}, we have 
\begin{equation}
Y_{n-1,2}=\frac{s\Gamma_{n,2}}{1-\Gamma_{n,2}}Y_{n-1,1}, 
\label{Laplaceeq}
\end{equation}
which reduces to 
\begin{equation}
Y_{n-1,2}=\frac{sf_{n,3}+f_{n,2}}{s-f_{n,2}}Y_{n-1,1}.
\end{equation}
We therefore have the headway to headway transfer function:
\begin{equation}
\Gamma_{n,1}=\frac{sf_{n3}+f_{n2}}{s^2+s(f_{n,3}-f_{n,1})+f_{n,2}},\label{tfunc02}
\end{equation}
which is the same as the speed to speed transfer function, see equation~\eqref{tfunc01}. 

The transfer functions $\Gamma_{n,k}$, with $k\in\{1,2\}$, have second order dynamics, therefore we can get an analytical condition for $\mathcal{L}_2$ strict string stability. The $\mathcal{H}_\infty$ norm of $\Gamma_{n,k}$ is the maximum gain $|\Gamma_{n,k}(j\omega)|$ across all frequencies. We have:
\begin{equation}
|\Gamma_{n,k}(j\omega)|=\sqrt{\frac{\omega^2f_{n,3}^2+f_{n,2}^2}{(f_{n,2}-\omega^2)^2+\omega^2(f_{n,3}-f_{n,1})^2}}.\label{amplitude}
\end{equation}
Condition $|\Gamma_{n,k}(j\omega)|\leq 1$ leads to equation
\begin{equation}
\omega^4+\omega^2(f_{n,1}^2-2f_{n,3}f_{n,1}-2f_{n,2})\geq 0.\label{condtf1}
\end{equation}
That is, the $\mathcal{L}_2$ strict string stability condition is a simple condition on the partial derivatives of the system: $\forall \omega \in \mathbb{R}^+$,
\begin{equation}
|\Gamma_{n,k}(j\omega)|\leq 1\Leftrightarrow f_{n,1}^2-2f_{n,1}f_{n,3}-2f_{n,2}\geq 0.\label{cond1}
\end{equation}  
Note that these conditions are the same conditions as the well-known string stability conditions derived for an infinite homogeneous traffic~\cite{wilson}. Note also that the discrepancy observed between the MIMO and SISO analysis may stem from the conservativeness of condition~\eqref{submu1}, and the fact that in the MIMO set-up the outputs are headway and speed and the input is just the speed of the previous vehicle.


\subsubsection{$\mathcal{L}_2$ weak string stability}{\label{helloparam}}
applying Definition 4, for a given $l\in \{i,...,m\}$ and $n\in\{l,...,m\}$, a sufficient condition for $(l,n)$ weak string stability is:
\begin{equation}
\norme{\prod_{i=l+1}^{n}\vv{\Gamma}_i}_{\mathcal{H}_\infty}\leq 1.\label{submu3}
\end{equation}

\begin{figure}[h!]
\includegraphics[width=\textwidth]{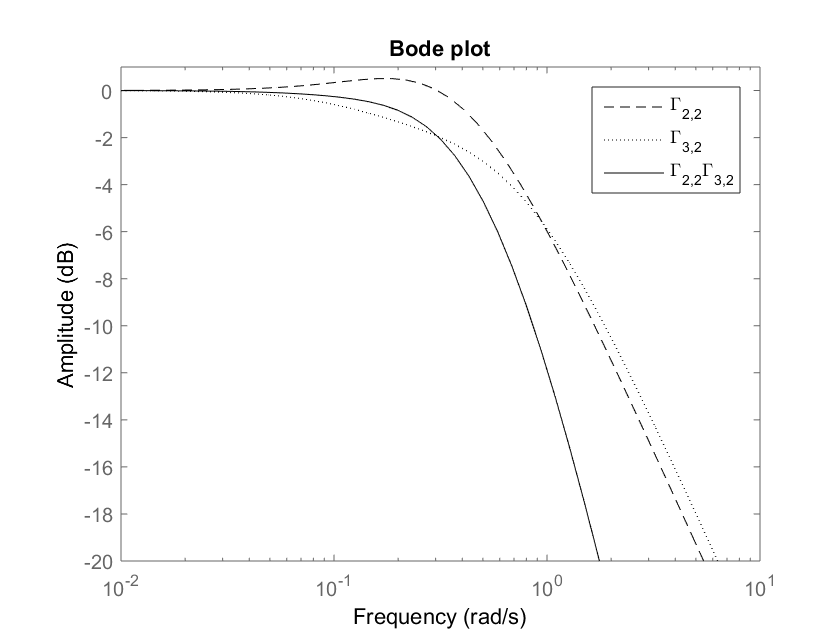}
\caption{System of 3 vehicles: Bode plots of $|\Gamma_{2,2}(j\omega)|$, $|\Gamma_{3,2}(j\omega)|$ and $|\Gamma_{2,2}(j\omega)\Gamma_{3,2}(j\omega)|$.}
\label{fig_1}
\end{figure}

As an example, consider a system consisting of 3 vehicles, with a disturbance on vehicle 1. We work with the following partial derivative values: $f_{21}=-0.075$, $f_{22}=0.091$, $f_{23}=0.55$, and $f_{31}=-0.26$, $f_{32}=0.10$, $f_{33}=0.64$, which correspond to realistic parameter values of the Intelligent Driver Model (IDM), a well-known physical model for reproducing realistic traffic~\cite{kesting1}. The $\mathcal{L}_2$ speed gains are $||\Gamma_{2,2}||_{\mathcal{H}_\infty}=1.06$, $||\Gamma_{3,2}||_{\mathcal{H}_\infty}=1$ and $||\Gamma_{2,2}\Gamma_{3,2}||_{\mathcal{H}_\infty}=1$, as we can observe in Figure~\ref{fig_1}. The system is therefore (1,3) weakly string stable but not proved to be strict string stable. The automated vehicle (here vehicle 3) could compensate instabilities generated by the conventional vehicle (here vehicle 2). However, analytical conditions for achieving weak string stability are not easy to obtain as solving inequality~\eqref{submu3} requires solving an $8$ degree polynomial equation.

\textit{Remark 5.} Note that, $\forall k\in\{1,2\}$, $\norme{\prod_{i=l+1}^{n}\Gamma_{i,k}}_{\mathcal{H}_\infty}\leq 1$ is equivalent to $\norme{\prod_{i=l+1}^{n}\Gamma_{i,k}}_{\mathcal{H}_\infty}= 1$, as $\forall i\in\{1,..,m\}$, we have $|\Gamma_{i,k}(0)|=1$.

\subsubsection{$\mathcal{L}_\infty$ strict string stability} as discussed in Section~\ref{3}, $\mathcal{L}_\infty$ string stability is more practical than $\mathcal{L}_2$ string stability as it deals with the peak values of the perturbations. One of the early works to introduce the $\mathcal{L}_\infty$-induced norm of a linear map, that is the $\mathcal{L}_1$ norm of its impulse response, are the ones of~\cite{vidyasagar1986optimal,dahleh19871}. This means that the condition to guarantee $\mathcal{L}_\infty$ strict string stability is to have the $\mathcal{L}_1$ norm of the impulse response less than 1. It is known from~\cite{boyd1991linear} that the $\mathcal{H}_{\infty}$ norm is upper bounded by the $\mathcal{L}_\infty$-induced norm, and that for non-negative impulse responses, those norms are identical. Therefore, if we look at the transfer functions of equations~\eqref{tfunc01},~\eqref{tfunc02}, necessary and sufficient conditions for having a monotonic step response are non-imaginary poles and negative zeros, which leads to:
\begin{align}
(f_{n,3}-f_{n,1})^2-4f_{n,2}&\geq 0,\label{tfunc2}\\
-\frac{f_{n,2}}{f_{n,3}}&<0.\label{tfunc3}
\end{align}
Note that the conditions~\eqref{tfunc3} is always verified due to the physical relation~\eqref{rat}. It is interesting to investigate which equation is the most conservative between~\eqref{cond1} and~\eqref{tfunc2}. In fact, by substracting equation~\eqref{cond1} from equation~\eqref{tfunc2}, we can verify that the $\mathcal{L}_{2}$ strict stability condition is stronger than the condition for the equality of the norms if $f_{n,3}^2\geq 2f_{n,2}$, which we expect to be almost always verified for realistic parameter values. In summary, we have:
\begin{equation}
f_{n,3}^2\geq 2f_{n,2} \Rightarrow \left(\mathcal{L}_{\infty}\ \text{stability}=\mathcal{L}_{2}\ \text{stability}\right),\label{l2linf}
\end{equation}
and if condition~\eqref{l2linf} is not verified we only have $\mathcal{L}_{\infty}\ \text{stability}\Rightarrow\mathcal{L}_{2}\ \text{stability}$. To our knowledge, whereas equation~\eqref{cond1} is well-known to the traffic flow theory community, equations~\eqref{tfunc02},~\eqref{tfunc2} and~\eqref{l2linf} are novel conditions for the investigated car-following dynamics of equations~\eqref{CF2} and~\eqref{CF1}.

\subsection{Closed vehicle systems}

We conclude the Section on stability results with the particular case of closed systems. Closed vehicle systems, where vehicle $1$ follows vehicle $m$, have often been studied to interpret congestion as they enable easy field experiments, see e.g.~\cite{sugiyama2008traffic}, and as it seems intuitive that the asymptotic instability of the closed system is linked to the string stability of the open system, see e.g.~\cite{wilson}. Here we present novel analytical results highlighting the difference between string and asymptotic stability in such systems.

By analogy with equation~\eqref{eq:matrice}, the system matrix $\vv{a_c}\in \mathbb{R}^{2(m+1)\times 2(m+1)}$ for a closed system can be written as
\begin{equation}
\vv{a_c}=\begin{pmatrix}
\vv{a}_{1,1}&0&\ldots&0&\vv{a}_{1,0}\\
\vv{a}_{2,0}&\ddots&\ddots&\ddots&0\\
0&\ddots&\ddots&\ddots&\vdots\\
\vdots&\ddots&\ddots&\ddots&0\\
0&\ldots&0&\vv{a}_{m,0}&\vv{a}_{m,1}
\end{pmatrix}.\label{eq:mAc2}
\end{equation}
For disturbance input $d_i$ at vehicle $i$, where $i>2$, we no longer have $\vv{y}_n(t)=0$ for $n<i$ since now the vehicles $n<i$ are affected by the disturbance on vehicle $i$. The dynamics of vehicle $i$ are now written as
\begin{equation}
\vv{Y}_i(s)=\vv{\Gamma}_i(s)\vv{Y}_{i-1}(s) + (s\vv{I}-\vv{a}_{i,1})^{-1}\vv{D}_i(s).
\end{equation}
In the case where the disturbance $d_1$ in on vehicle $1$, we have $\vv{Y}_1(s)=\vv{\Gamma}_1(s)\vv{Y}_{m}(s) + (s\vv{I}-\vv{a}_{1,1})^{-1}\vv{D}_1(s)$, where $\vv{\Gamma}_1(s)$ is the transfer function for vehicle $m$ to vehicle $1$. We then have
\begin{equation}
\vv{Y}_i(s)=\prod_{j=1}^m\vv{\Gamma}_j(s)\vv{Y}_{j}(s) + (s\vv{I}-\vv{a}_{i,1})^{-1}\vv{D}_i(s),
\end{equation}
and finally, with $d_i$, $\forall i\in \{1,...,m\}$, being the only non-zero component, we have
\begin{equation}
\vv{Y}_i(s)=\left(\vv{I}-\prod_{j=1}^m\vv{\Gamma}_j(s)\right)^{-1}\vv{G}_i(s)\vv{D}_i(s).
\label{asy-str}
\end{equation}

\subsubsection{SISO homogeneous case} we investigate the particular case where $d_i\in\mathbb{R}$, $i\geq 1$, $y_n\in\mathbb{R}$, $\forall n\in\{1,\ldots,m\}$, $\forall k\in\{1,2\}$, $\Gamma_{n,k}=\Gamma_{1}$. Following Remark 1 and equation~\eqref{asy-str}, as $(a_{n,1})$ is a Hurwitz matrix, exponential stability is achieved when the poles of $(1-\Gamma_1^n)^{-1}$ have negative real parts. We factorise $(1-\Gamma_1^n)$ as
\begin{equation}
1-\Gamma_1^n=(1-\Gamma_1)\prod_{k=1}^{m-1}\left(\Gamma_1-e^{\frac{2ik\pi}{m}}\right),
\end{equation}
and developing from equation~\eqref{tfunc01}, the denominator $D_c$ of $(1-\Gamma_1^n)^{-1}$ is equal to
\begin{small}
\begin{equation}
D_c=\prod_{k=0}^{m-1}\left(s^2-s\left(f_1+f_3\left(e^{\frac{2ik\pi}{m}}-1\right)\right)-f_2\left(e^{\frac{2ik\pi}{m}}-1\right)\right)\label{rootssiso}.
\end{equation}
\end{small}
Note that this expression closely resembles the condition for string stability in the infinite homogeneous case derived using the Fourier perturbation technique~\cite{wilson}. The infinite homogeneous system is said to be stable iff $\forall k\in[0,2\pi],\ s^2-s\left(f_1+f_3\left(e^{ik}-1\right)\right)-f_2\left(e^{ik}-1\right)$ has negative real parts. This condition can be shown to be equivalent to the $\mathcal{L}_{2}$ strict stability condition of equation~\eqref{cond1}, see~\cite{monteil2}.

\subsubsection{General case} in the general case, exponential stability is achieved when the transfer function in equation~\eqref{asy-str} has poles with negative real parts. From equation~\eqref{weakstr}, we have 
\begin{equation}
\prod_{j=1}^m\vv{\Gamma}_j(s)=\frac{1}{\prod_{j=1}^m\gamma_j(s)}\begin{pmatrix}
0&P_1(s)\\
0&P_2(s)
\end{pmatrix},
\end{equation}
where $\gamma_i(s)=s^2+s(f_{j,3}-f_{j,1})+f_{j,2}$, and $P_1(s)$ and $P_2(s)$ are polynomials of degree $m$, with 
\begin{equation}
P_2(s)=\prod_{i=1}^m(f_{i2}+sf_{i3}).
\end{equation} 
Given that $\forall i \in \{1,...,m\}$, the solutions of $\gamma_i(s)=0$ have negative real parts, the system is exponentially stable iff the solutions of the following equation
\begin{equation}
\prod_{i=1}^m\gamma_i(s)-P_2(s)=0,
\end{equation}
have negative real parts. 


\textit{Remark 6}.
There are situations for which the closed vehicle system is asymptotically stable but not strict string stable, i.e. equation~\eqref{cond1} is not verified. For example, for 3 homogeneous vehicles, choosing $f_{n,1}=-0.075$, $f_{n,2}=0.091$, $f_{n,3}=0.55$ as in Section~\ref{helloparam}, with $n \in\{1,...,3\}$, we have $||\Gamma_n||_{\mathcal{H}_\infty}>1$ while the eigenvalues of matrix $\vv{a_c}$ have negative real parts.

\section{Simulation}\label{5}

In this section we illustrate the previous analytical results regarding string stability using simulations. 

\subsection{Model selection and parameter distributions}\label{parameters}

The Intelligent Driver Model (IDM)~\cite{kesting2010enhanced} defines function $f_n$ of equation~\eqref{CF1} as: 
\begin{align}
\label{IDM}
f_n(\dot{x}_n,\Delta x_n, \Delta \dot x_n)=a\left[1-\left(\frac{\dot x_n}{V_{\max,n}}\right)^{4}-\left(\frac{s^{\star}(\dot{x}_n,\Delta \dot x_n)}{\Delta x_n-l_n}\right)^2\right],
\end{align}
where 
\begin{equation}
s^{\star}(\dot{x}_n,\Delta \dot x_n)=s_{0,n}+\max\left(0,\dot x_n T_n-\frac{\dot x_n \Delta \dot x_n}{2\sqrt{a_nb_n}}\right),
\end{equation}
where the following behavioural parameters are specific to vehicle $n$: $V_{\max,n}$ is the desired free-flow speed, $T_n$ is the safe time headway, $a_n$ is the maximum tolerated acceleration, $b_n$ is the comfortable deceleration, and $s_{0,n}$ is the minimum stopping distance. 

In order to reproduce realistic heterogeneous traffic in simulation, we have developed a complete methodology to perform robust offline parameter identification starting from noisy trajectory data~\cite{prac}, which involves sensitivity analysis, point estimation and interval estimation. The parameter estimates we use here are the outputs of this methodology, for the 3 most left lanes of the well-known US101 NGSIM dataset~\cite{NGSIM}, during morning peak time (7:50am to 8:05am). The estimates were found to fit log-normal distributions for parameters $a$ and $b$ and normal distributions for parameters $T$ and $s_0$. The mean and standard deviations are $m_a=0.77~\si{m.s^{-2}}$, $\sigma_a=0.42~\si{m.s^{-2}}$, $m_b=1.1~\si{m.s^{-2}}$, $\sigma_b=0.43~\si{m.s^{-2}}$, $m_T=1.5~\si{s}$, $\sigma_T=0.57~\si{s}$, $m_{s_0}=2~\si{m}$, and $\sigma_{s_0}=0.5~\si{m}$. To reproduce heterogenous traffic, the parameters are sampled from these distributions truncated at the physical bounds chosen as in the literature: $a\in [0.3,3]$, $b\in [0.3,3]$, $T\in [0.3,3]$, $s_0\in [0.5,3.5]$~\cite{punzo2015we}. With $V_{\max}=33~\si{m.s^{-1}}$ roughly corresponding to the speed limit of the section, see~\cite{punzo2015we}, and taking for instance an equilibrium speed of $V_{\text{eq}}=V_{\max}/2$, this gives for an average vehicle $n$ $f_{n,1}^2-2f_{n,1}f_{n,3}-2f_{n,2}=-0.063$, meaning that parameters are distributed so that the $\mathcal{L}_{2}$ string stability condition is not verified for a number of vehicles. In the remainder of the paper we write $S_n:=f_{n1}^2-2f_{n,1}f_{n,3}-2f_{n,2}$.

\subsection{Homogeneous traffic: $\mathcal{L}_2$ strict string stability}

We focus on homogeneous traffic first, i.e. when the behavioural parameters are the same for all vehicles.

\subsubsection{Relevance of the strict string stability condition}\label{relevancess} since the automated vehicles obey the IDM car-following model, it is of interest to investigate the parameter space for which the model exhibits strict and weak string stable behaviour. In this subsection we focus on the evolution of the two most sensitive parameters, parameters $a$ and $T$, see~\cite{punzo2015we}, to gain insights on the possibilities to reach string stable behaviour with realistic parameters values. The rest of the parameters are chosen to have realistic values, see Section~\ref{parameters}, i.e. $b=1.1~\si{m.s^{-2}}$, $s_0=2~\si{m}$ and $V_{\max}=33~\si{m.s^{-1}}$.

\begin{figure}[h!]
\begin{tabular}{cc} 
\subfloat[]{\includegraphics[width=0.5\textwidth]{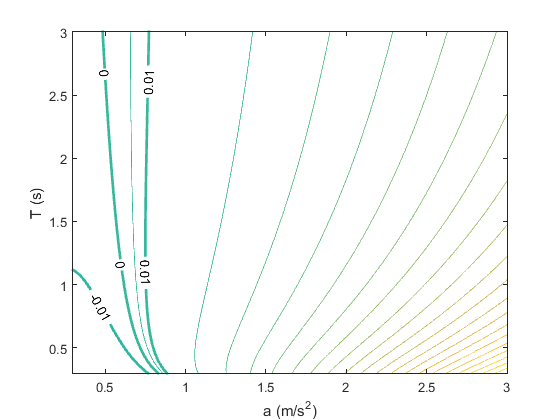}\label{fig_C1}}&
\subfloat[]{\includegraphics[width=0.5\textwidth]{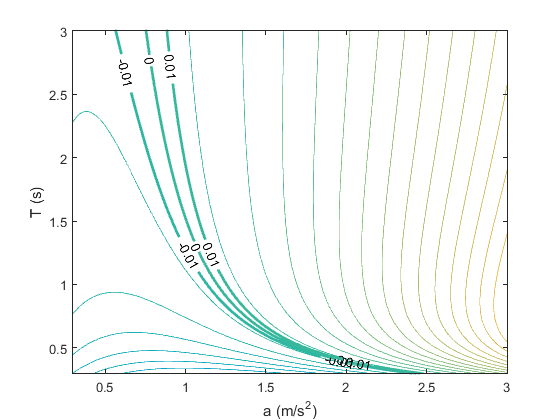}\label{fig_C3}} 
\end{tabular}
\caption{Contour lines of the string stability coefficient $S_n$ for (a) $V_{eq}=2V_{\max}/3$, (b) $V_{eq}=V_{\max}/3$.}
\label{Contour}
\end{figure}

Figure~\ref{Contour} plots the contour lines of the string stability coefficient $S_n$. It can be seen that the limit between string stability and string instability depends on the traffic equilibrium speed. For low equilibrium speeds of $V_{\max}/3$, which roughly corresponds to the value observed for the NGSIM data set, we approximately need $a\geq 1.1~\si{m.s^{-2}}$ with $T\geq 1.6~\si{s}$ to have a string stable system (positive $S_n$). The instability domain is increased as we move towards lower equilibrium speeds. 

\subsubsection{Strictly string stable vs strictly string unstable traffic} for the previously defined parameters, with $T=1.5~\si{s}$, $V_{\text{eq}}=V_{\max}/2$, consider a string unstable system with $a=0.47~\si{m.s^{-2}}$, and so $S_n=-0.018<0$; and a string stable system for $a=0.87~\si{m.s^{-2}}$, and so $S_n>0$. A disturbance is introduced on vehicle 1 in the form of a unit step of $-1~\si{m.s^{-2}}$ between $t_1=5~\si{s}$ and $t_2=10~\si{s}$. Note that this actually corresponds to the sum of two opposed input steps, one happening at $t_1=5~\si{s}$ and one at $t_2=10~\si{s}$.

\begin{figure}[h!]
\begin{tabular}{cc} 
\subfloat[]{\includegraphics[width=0.5\textwidth]{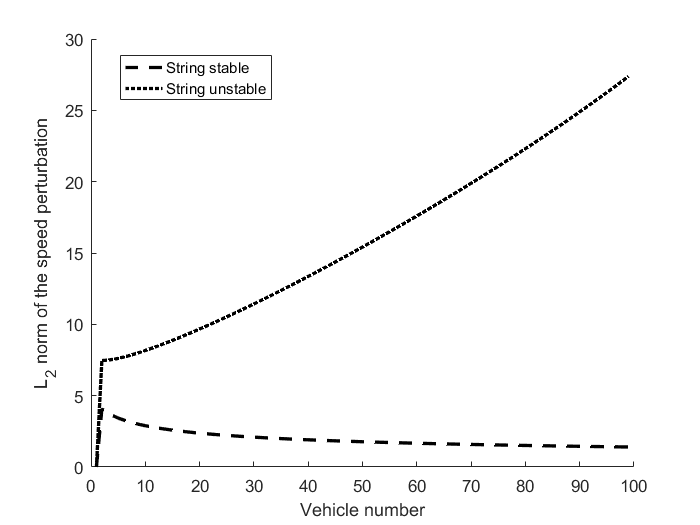}\label{fig_s1}}&
\subfloat[]{\includegraphics[width=0.5\textwidth]{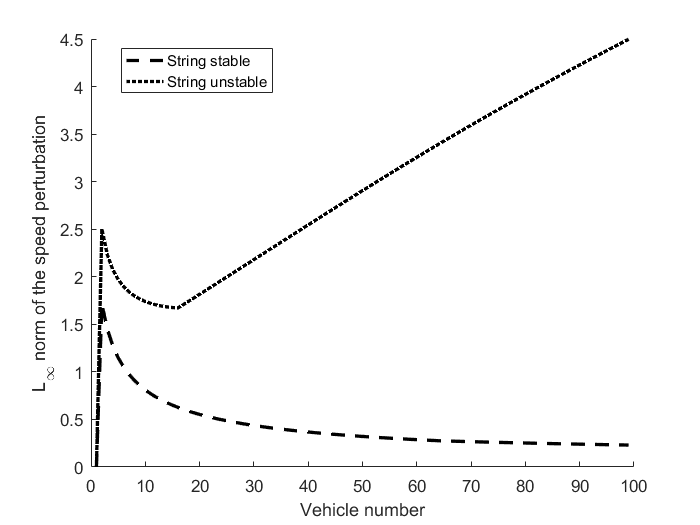}\label{fig_s2}}
\end{tabular}
\caption{Evolution of the speed perturbations under a disturbance of $A=-1~\si{m.s^{-2}}$ as function of the vehicle number for string stable and string unstable systems: (a) $\mathcal{L}_2$ norm; (b) $\mathcal{L}_{\infty}$ norm.}
\label{L2normpos}
\end{figure} 

\begin{figure}[h!]
\begin{tabular}{cc} 
\subfloat[]{\includegraphics[width=0.5\textwidth]{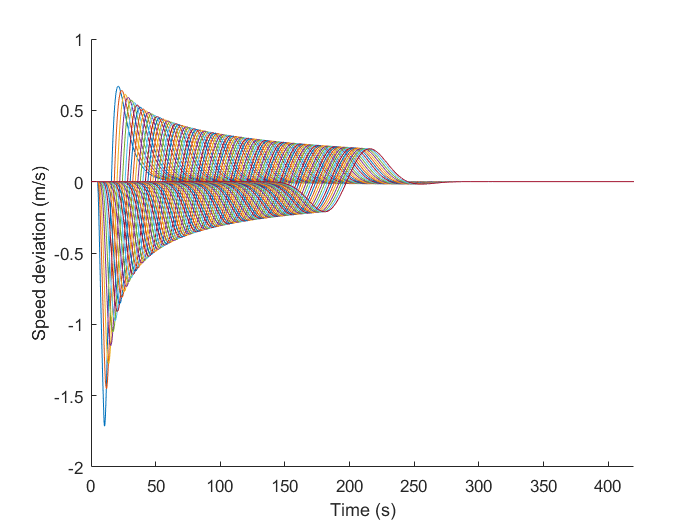}\label{fig_s3}\label{L2posn1}}&
\subfloat[]{\includegraphics[width=0.5\textwidth]{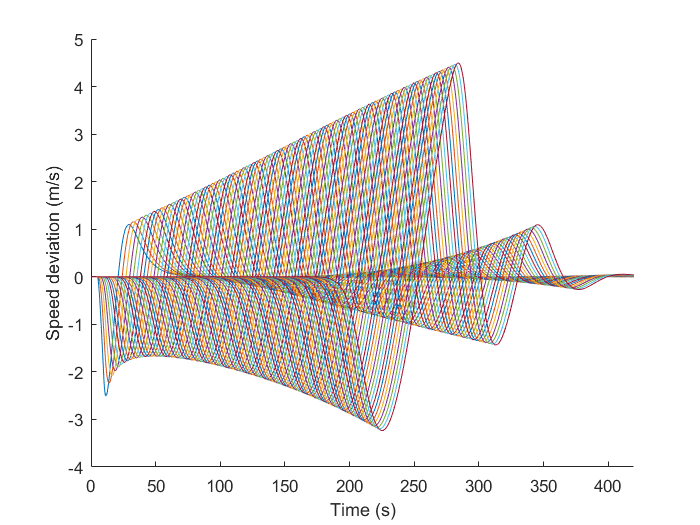}\label{fig_s4}\label{L2posn2}}
\end{tabular}
\caption{Evolution of the speed perturbation following a disturbance of $A=-1~\si{m.s^{-2}}$ for (a) a strictly string stable system, (b) a strictly string unstable system.}
\label{L2normneg}
\end{figure}

The $\mathcal{L}_2$ norms of the speed perturbations are computed using an Euler sum over the simulation time steps. It can be seen from Figure~\ref{L2normpos} that, in the strictly string stable case, the $\mathcal{L}_2$ and $\mathcal{L}_\infty$ norms monotonically decrease with the vehicle number. Conversely, in the strictly string unstable case, it can be seen that while the $\mathcal{L}_{\infty}$ norm initially decreases, both norms tend to increase after a certain vehicle number is reached. This is in accordance with the conclusions of Section~\ref{strictweak}. Figure~\ref{L2normneg} displays the evolution of the speed perturbation for all the vehicles in both strictly string stable and strictly string unstable cases. The string stability property means that the perturbation fades away. Note that we could have focused on the evolution of the headway perturbation equivalently as it leads to similar observations, as per equations~\eqref{tfunc01},~\eqref{tfunc02}. 

A last remark is made in the light of Figures~\ref{L2normpos} and~\ref{L2normneg}. It is observed that the perturbation does not completely vanish, i.e. the bounded disturbance is not attenuated to a perfect zero $\mathcal{L}_2$ norm as we move downstream. This is related to the fact that $||\Gamma_{n,2}||_{\mathcal{H}_{\infty}}$ asymptotically converges towards $\Gamma_{n,2}(0)=1$, see equation~\eqref{cond1} and Figure~\ref{fig_1}, and that~\eqref{submu1} is not a strict inequality, which means that the strict string stability condition does not require long-wave perturbations to be attenuated at a specific rate. Therefore, a stronger condition than $\mathcal{L}_2$ strict string stability would be to force a sharper decrease of the Bode plot for low frequencies, see Figure~\ref{fig_1}.



\begin{figure}[h!]
\begin{tabular}{cc} 
\subfloat[]{\includegraphics[width=0.5\textwidth]{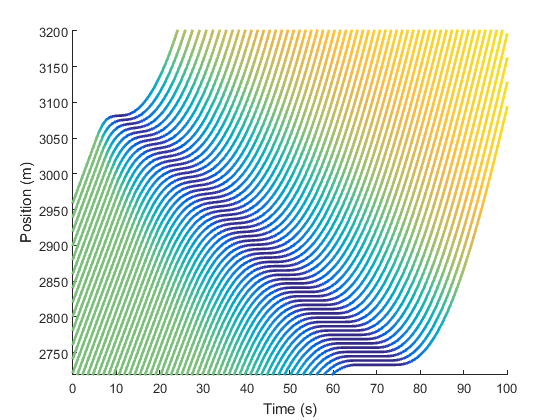}\label{fig_s3}\label{fig_NL1}}&
\subfloat[]{\includegraphics[width=0.5\textwidth]{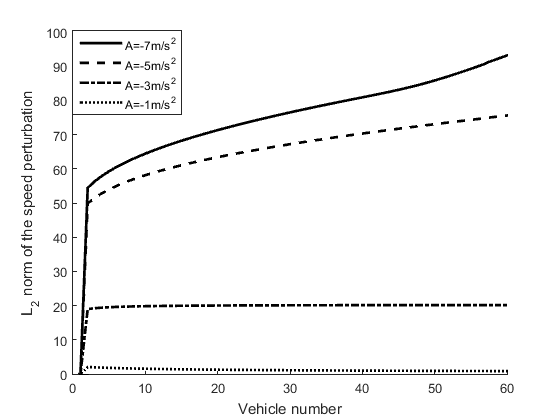}\label{fig_s4}\label{fig_NL2}}
\end{tabular}
\caption{Evolution of the (a) time vs position diagram for a string stable system following a disturbance of $A=-7~\si{m.s^{-2}}$; (b) $\mathcal{L}_2$ norm of the speed perturbation for a string stable system and varying accelerations inputs.}
\label{fig_NL3}
\end{figure}

\subsubsection{Nonlinear vs linear string stability: empirical observations}\label{lnnonln}

let us now briefly discuss the dependence of string stability on the size of the disturbance. For the situation where $a=1.55~\si{m.s^{-2}}$, and $T=0.8~\si{s}$, $b=1.7~\si{m.s^{-2}}$, we have $S_n=0.0038>0$. Figure~\ref{fig_NL1} presents the evolution of the time-position diagram for a disturbance of $-7~\si{m.s^{-2}}$ between $t_1$ and $t_2$ and Figure~\ref{fig_NL2} presents the evolution of the $\mathcal{L}_2$ norm of the speed perturbation for varying disturbances. 

It can be seen that, for disturbances of $-5~\si{m.s^{-2}}$ and $-7~\si{m.s^{-2}}$, the values of the $\mathcal{L}_2$ norm of the speed perturbation are growing as we move downstream the vehicle system. This contradicts the string stability condition~\eqref{cond1}. We see that, for a disturbance of $-7~\si{m.s^{-2}}$, the perturbation is being amplified until the vehicles completely stop, and the $\mathcal{L}_2$ norm seems to be unbounded. For other disturbances, Figure~\ref{fig_NL2} shows that the slope of the $\mathcal{L}_2$ norm curve seems to be decreasing as we move downstream the vehicle system. Finally, when the intensity of the disturbance is kept within realistic values, i.e. $A<-5~\si{m.s^{-2}}$, the $\mathcal{L}_2$ norms appear to remain bounded as the number of vehicles in the system is increased.

Such observations are related to nonlinear effects and to the non-validity of the linearisation hypothesis. If the low speed area spreads through the time-space, the linearised dynamics which satisfy the string stability condition is not valid anymore, and another linearisation about a lower equilibrium speed would indicate string instability, as Figure~\ref{Contour} suggests. Besides, the car-following formulation we consider does not deal with the zero speed constraint, i.e. the fact that vehicles cannot have negative speeds.

\subsection{Heterogeneous traffic: $\mathcal{L}_2$ weak strict string stability}

In this section we present an example that highlights the relevance of verifying the weak string stability condition, see equation~\eqref{submu3}. We consider a system composed of 30 vehicles having different behaviour, and we introduce a disturbance of $-1~\si{m.s^{-2}}$ on vehicle 1 between $t_1$ and $t_2$. The variability in the vehicle system is introduced by sampling parameters $a$ and $T$ from truncated distributions as described in Section~\ref{parameters}. For instance, we can get a $(0,30)$ weakly string stable system, i.e. $\prod_{i=1}^{30}||\Gamma_{i,2}||_{\mathcal{H}_{\infty}}\leq 1$, when $a$ and $T$ are sampled from the distributions presented in Section~\ref{parameters} but defining $a\in[0.5,3]$ and $T\in[1.1,3]$; and we can get a $(0,30)$ weakly string unstable system, i.e. $\prod_{i=1}^{30}||\Gamma_{i,2}||_{\mathcal{H}_{\infty}}=1.94>1$, by defining $a\in[0.3,1]$ and $T\in[0.3,2]$. Note that the $(0,30)$ weakly string stable and weakly string unstable systems are obtained for particular samples of the truncated distributions.

\begin{figure}[h!]
\begin{tabular}{cc} 
\subfloat[]{\includegraphics[width=0.5\textwidth]{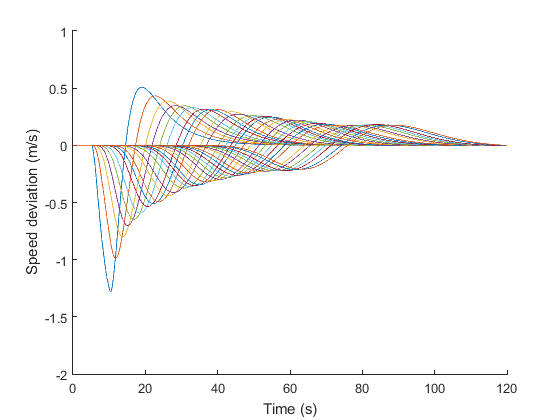}\label{fig_WS1}}&
\subfloat[]{\includegraphics[width=0.5\textwidth]{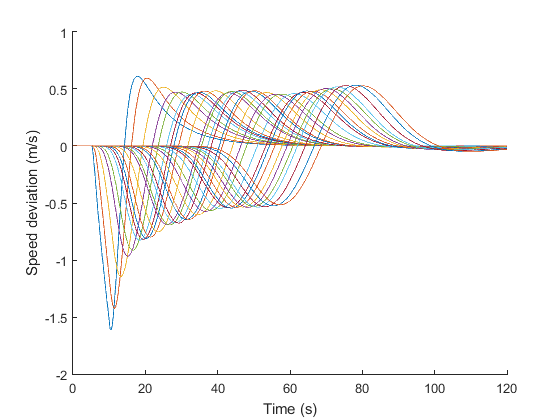}\label{fig_WS2}}
\end{tabular}
\caption{Evolution of the speed perturbation for successive vehicles in the case of: (a) $(0,30)$ weak string stability; (b) $(0,30)$ weak string instability.}
\label{fig_WS3}
\end{figure} 

Figure~\ref{fig_WS3} shows the evolution of the speed perturbations within the vehicle system. It appears that the speed deviation is being damped in the weakly string stable case, despite the presence of 8 strictly string unstable vehicles in the considered vehicle system, and is being amplified in the weakly string unstable case, despite the presence of 9 strictly string stable vehicles. 

In the remainder of this paper, we will investigate how to tune the behavioural parameters of the automated vehicles so as to increase the weak string stability of the traffic flow. 

\section{Parameter optimisation}\label{6}

In this section, the automated vehicles update their longitudinal dynamics according to the IDM car-following model. We formulate the following optimisation problem: the vehicle behavioural parameters of each automated vehicle are picked to maximise the strict/weak string stability of the system, while minimising the distance between their parameter values and the vehicle behavioural parameters when there is no control, e.g. in the case of a partially automated vehicle with automated and non-automated driving modes, when the automated mode is deactivated. Note that we have made available a simple example of the code at~\href{https://uk.mathworks.com/matlabcentral/fileexchange/62774-toy-example-string-stability}{this hyperlink}.

\textit{Remark 7.} Note that we make the choice to rely on in-vehicle sensors to estimate the vehicle behavioural parameters, and to use those estimated parameters to design the optimisation policy so as to increase string stability. Another way of increasing string stability is to utilise the car-following model structure itself to integrate V2V communication, see~\cite{monteil2,Ngoduy2015420} for instance, however by doing that the safe structure of car-following dynamics is lost, i.e. collisions may occur. Our approach of optimising the car-following parameters is key to preserving the safe structure of the car-following dynamics. This enables the design of safe ACC systems that takes into consideration the driving behaviour of the surrounding vehicles in heterogeneous traffic as well as the driving comfort of the driver in the automated vehicle.


\subsection{Generic formulation of the optimisation problem}\label{sstpolicy}

Let $\mathcal{T}$ denote the joint distribution of the car-following parameters. The vector of parameters $\theta_i\in\mathbb{R}^l$, $i\in\{1,...,m\}$, defining the dynamics of each vehicle $i$ is sampled from this distribution. We write the covariance matrix of $\mathcal{T}$ as $\vv{\Sigma_{\mathcal{T}}}\in\mathbb{R}^{k\times k}(\mathbb{R})$. When there is no correlation between parameters, as in Section~\ref{parameters}, $\Sigma_{\mathcal{T}}$ is diagonal, and the elements of the diagonal are the inverses of the standard deviations of each parameter. For each automated vehicle indexed $n$, we seek to optimise the $k\in\{1,...,l\}$ parameters $\theta_n$, with $\Theta\subset\mathbb{R}^k$ denoting the admissible set of parameter values, e.g. the physical bounds of the parameters defined in section~\ref{parameters}. Therefore we have $\theta_n\in\Theta$. In the case of a partially-automated vehicle, $\hat{\theta}_{n}$ denotes the estimated behavioural parameters for vehicle $n$ when the automated mode is deactivated; in the case of a fully-automated vehicle, $\hat{\theta}_{n}$ designates average comfortable driving parameters. 

\subsubsection{Relaxation of weak string stability}\label{relaxweak}

in Section~\ref{helloparam}, we discussed how automated vehicles can be used to achieve weak string stability. However, there exist situations for which this is not possible: for example, consider a system of 3 vehicles with parameters $a_1=0.58~\si{m.s^{-2}}$, $a_2=0.35~\si{m.s^{-2}}$, $a_3=0.39~\si{m.s^{-2}}$, and $T_1=1.76~\si{s}$, $T_2=1.26~\si{s}$, $T_3=1.43~\si{s}$. The rest of the parameters are chosen as $V_{\max}=33~\si{m.s^{-1}}$, $b=1.1~\si{m.s^{-2}}$, and $s_0=2~\si{m}$ as in Section~\ref{parameters}, and $V_{eq}=V_{\max}/3$. For this situation, we have $\prod_{i=1}^{3}||\Gamma_{i,2}||_{\mathcal{H}_{\infty}}=1.12>1$, and after numerical simulations we find no values of $a_4\in[0.3,3]$ and $T_4\in[0.3,3]$ leading to $\prod_{i=1}^{4}||\Gamma_{i,2}||_{\mathcal{H}_{\infty}}=1$. This means that the weak string stability constraint of equation~\eqref{submu3} can not be used as a hard constraint to an optimisation policy. Consequently, in the next sections we relax constraint $\prod_{i=1}^{4}||\Gamma_{i,2}||_{\mathcal{H}_{\infty}}=1$ to $\prod_{i=1}^{4}||\Gamma_{i,2}||_{\mathcal{H}_{\infty}}<\gamma$, with $\gamma\in\mathbb{R}, \ \gamma\geq 1$. 


\subsubsection{Optimisation problem}\label{optimp} let $i$ and $j$ be the farthest upstream and downstream vehicles for which parameter estimates $\hat{\theta}_i$ and $\hat{\theta}_j$ are known. We have $1\leq i\leq n\leq j\leq m$. If there is no knowledge of the behaviour of upstream and downstream vehicles, then $i=j=n$. Constraints are placed on the $\mathcal{L}_2$ gain between the speed perturbation of vehicle $i-1$ and the speed perturbation of vehicle $j$, i.e reflecting our aim of achieving $(i-1,j)$ weak string instability, see equation~\eqref{submu3}. The decision variables are the behavioural parameters $\theta_n$ of the partially-automated vehicle $n$.

We propose the following optimisation problem to capture these design requirements:
\begin{align}
&\min_{\theta_n,\gamma}\alpha\gamma+\frac{1}{k}\left(\theta_{n}-\hat{\theta}_{n}\right)\vv{\Sigma}_{\mathcal{T}}^{-1}\left(\theta_{n}-\hat{\theta}_{n}\right)^T,\label{sstr1}\\
&\text{s.t.}\ \left\{
\begin{array}{l}
  \theta_n\in \Theta,\\
  \forall (i,j)\in \mathcal{N}_{n},\\
		||\Gamma_{i,2}\cdots\Gamma_{j,2}||_{\mathcal{H}_\infty}\leq \gamma
\end{array}
\right.
\label{sstr2}
\end{align}
where the objective is to minimise the distance between the optimised parameters $\theta_{n}$ and the vector of parameters of the vehicle $\hat{\theta}_{n}$ when the automation mode is deactivated, as well as to minimise $\gamma$. Constant $\alpha \in\mathbb{R_+^*}$ is a design parameter. $\mathcal{N}_{n}$ designates the set of pairs of neighbouring upstream and downstream vehicles for which parameter estimates $(\hat{\theta}_i)_{i\in \mathcal{N}_{n}}$ are known, with $i\leq n\leq j$. The $(i-1,j)$ weak string stability condition is relaxed as discussed in Section~\ref{relaxweak}.

\textit{Remark 8.} Note that the minimisation of the $\mathcal{H}_\infty$ norm of the input-output transfer function $||G_{i,2}\Gamma_{i+1,2}\cdots\Gamma_{j,2}||_{\mathcal{H}_\infty}$ could be formulated as another constraint. 

\textit{Remark 9.} Regarding the values of $i$ and $j$, in practice, automated vehicles are equipped with sensors which can enable parameter estimation for only a few leading/following vehicles. Looking at equation~\eqref{CF1}, the acceleration and speed of vehicle $n$, and the relative positions and speeds between vehicle $n$ and vehicle $n-1$ need to be tracked to be able to estimate $\theta_n$ via static parameter estimation techniques, see e.g.~\cite{monteil2015real}. Considering that only the positions and speeds of 2 upstream and downstream vehicles can be tracked with in-vehicle sensors, we rarely have $i<n-1$ and $j>n+2$ unless behavioural parameter data is transmitted via communication channels. 

\subsubsection{Limitations of weak string stability}\label{limitation} when the knowledge of behavioural parameters is limited to only a few leaders and followers, there exist situations for which the $(i-1,j)$ weak string stability constraint is verified but the $(i-1-i_1,j+j_1)$ weak string stability constraint is not, for given $i_1,j_1\in\mathbb{N}$. For example, taking $i=j=n$, we have $||\Gamma_{n,2}||_{\mathcal{H}_\infty}=1$ and $||\Gamma_{n,2}\Gamma_{n-1,2}||_{\mathcal{H}_\infty}>1$ for the following parameter values: $a_n=0.9~\si{m.s^{-2}}$, $b_n=0.9~\si{m.s^{-2}}$, $T_n=2.5~\si{s}$, $a_{n-1}=0.5~\si{m.s^{-2}}$, $b_{n-1}=1.7~\si{m.s^{-2}}$, $T_{n-1}=0.8~\si{s}$, with $V_{\max}=33~\si{m.s^{-1}}$, $s_0=2~\si{m}$, and $V_{\text{eq}}=V_{\max}/3$. This means that the verification of the $(i-1,j)$ weak string stability is not sufficient to ensure a $(0,m)$ weakly string stable system. However, there exist two ways to address this issue in order to provide a more stable system dynamics. The first one consists in considering the minimisation of the input-output $\mathcal{L}_2$ gain as well, that is the minimisation of $\norme{G_{n,2}}_{\mathcal{H}_{\infty}}$. The second one consists in adding one (or various) fictitious unstable leading or following vehicle(s), i.e worst case vehicle(s), which will eventually lead to more extreme parameter values compensating the fictitious instabilities. For instance, let us consider the case where the parameters of only vehicle $n$ and vehicle $n-1$ are known. Then, introducing a worst case vehicle with parameters $\theta_{wc}$, we now perform the minimisation of $||\Gamma_{wc,2}\Gamma_{n,2}\Gamma_{n-1,2}||_{\mathcal{H}_\infty}$.

\subsection{LMI formulation of the optimisation problem}\label{sstpolicy}

We can rewrite constraints~\eqref{sstr2} as Linear Matrix Inequalities~\cite{BEFB:94}. Starting from equation~\eqref{eq:sty2} and combining the linearised dynamics of cars $i$,$...$,$j$, with $i,j\in\mathcal{N}_{n}$ and $i\leq j$, we write the following linearised system dynamics:
\begin{align}
\dot{\vv{y}}_{i,j}&=\vv{A}_{i,j}\vv{y}_{i,j}+\vv{b}_{i,j}\vv{u}_{i,j},\label{DefLMI}\\
\vv{h}_{i,j}&=\vv{c}_{i,j}\vv{y}_{i,j}
\end{align}
where $\vv{y}_{i,j}^T=[\vv{y}_i^T,\vv{y}_{i+1}^T,\cdots,\vv{y}_{j}^T]$, $\vv{u}_{i,j}^T=[\vv{u}_i^T,\vv{u}_{i+1}^T,\cdots,\vv{u}_{j}^T]$, $\vv{b}_{i,j}\in\mathbb{R}^{2(j-i+1)\times 2(j-i+1)}$ and $\vv{c}_{i,j}\in\mathbb{R}^{2(j-i+1)\times 2(j-i+1)}$ are the input weighting and observation matrices, and 
\begin{align}
\vv{A}_{i,j}&=\begin{pmatrix}
\vv{a}_{i,1}&0&\ldots&\ldots&0\\
\vv{a}_{i+1,0}&\ddots&\ddots&\ddots&\vdots\\
0&\ddots&\ddots&\ddots&\vdots\\
\vdots&\ddots&\ddots&\ddots&0\\
0&\ldots&0&\vv{a}_{j,0}&\vv{a}_{j,1}
\end{pmatrix}.\label{eq:LMI}
\end{align}



As the speed $\dot{y}_j$ is observed, we have

\begin{equation}
\vv{c}_{i,j}=\begin{pmatrix}
0&\ldots&\ldots&\ldots&0\\
\vdots&\ddots&\ddots&\ddots&\vdots\\
\vdots&\ddots&\ddots&\ddots&\vdots\\
\vdots&\ddots&\ddots&0&0\\
0&\ldots&\ldots&0&1
\end{pmatrix}.
\end{equation}
The stability constraint is on the $\mathcal{L}_2$ gain between speed perturbation $\dot{y}_{i-1}$ and speed perturbation $\dot{y}_{j}$. We consider the input as $\vv{y}_{i-1}$. Following equation~\eqref{eq:sty2}, since the first column of matrix $\vv{a}_{i,0}$ consists of zeros, only $\dot{y}_{i-1}$ acts as input to the $\vv{y}_j$ dynamics, i.e. the $\Delta y_{i-1}$ term has no effect since it is multiplied by the zeros in the first column of $a_{i,0}$, which makes it a SISO system. We can therefore write
 
\begin{equation}
\vv{b}_{i,j,1}=\begin{pmatrix}
0&1&0&\ldots&0\\
\vdots&f_{i,3}&0&\ddots&\vdots\\
\vdots&\ddots&0&\ddots&\vdots\\
\vdots&\ddots&\ddots&\ddots&\vdots\\
0&\ldots&\ldots&\ldots&0
\end{pmatrix}.
\end{equation}

Therefore, using the LMI characterisation of the $\mathcal{L}_2$ gain, see~\cite{BEFB:94,isidori}, we can reformulate the optimisation problem~\eqref{sstr1},~\eqref{sstr2} as:
\begin{align}
&\min_{\theta_n,\vv{X}_{i,j},\gamma}\alpha\gamma+\frac{1}{k}\left(\theta_{n}-\hat{\theta}_{n}\right)\vv{\Sigma_{\mathcal{T}}}^{-1}\left(\theta_{n}-\hat{\theta}_{n}\right)\\
&\text{s.t.}\ \left\{
\begin{array}{l} 
  \theta_n\in \Theta,\\
\forall (i,j)\in \mathcal{N}_{n},\\  
\begin{pmatrix}
\vv{A}_{i,j}^T\vv{X}_{i,j}+\vv{X}_{i,j}\vv{A}_{i,j}&\vv{X}_{i,j}\vv{b}_{i,j,1}&\vv{c}_{i,j}^T\\
\vv{b}_{i,j,1}^T\vv{X}_{i,j}&-\gamma\vv{I}_{i,j}&\vv{0}\\
\vv{c}_{i,j}&\vv{0}&-\gamma\vv{I}_{i,j}
\end{pmatrix}
\prec 0,\\
\vv{X}_{i,j}\succ 0,
\end{array}
\right.\label{LMIopt2}
\end{align}
where the matrices $\vv{A}_{i,j}$ depend on $\theta_n$, see equations~\eqref{eq:a0},~\eqref{eq:a1},~\eqref{eq:LMI}, $\vv{X}_{i,j}\in\mathbb{R}^{2(j-i+1)\times 2(j-i+1)}$, and $\vv{I}_{i,j}\in\mathbb{R}^{2(j-i+1)\times 2(j-i+1)}$ is the identity matrix. This optimisation problem is convex in $\vv{X}_{i,j}$, $\gamma$, but not in the parameters of the car-following model $\theta_n$, and not jointly convex in $\vv{A}_{ij}$ and $\vv{X}_{ij}$. Even after linearising or convexifying the car-following model, assuming $\vv{\Sigma_{\mathcal{T}}}^{-1}$ to be positive semi-definite, we would still be facing a biconvex optimisation problem. In this paper, we explore the car-following model parameter space using simulated annealing and solve the convex part of the optimisation using cvx~\cite{cvx} to obtain $\vv{X}_{i,j}$ and $\gamma$ at each iteration. Note that other heuristics such as the Alternate Convex Search (ACS)~\cite{gorski2007biconvex} may be of use. 

\textit{Remark 10.} The constraint concerning the minimisation of the $\mathcal{L}_2$ gain between the disturbance $d_{i}$ and the speed perturbation $\dot{y}_{j}$, mentioned in Remark 8, can also be formulated as LMI.

%
%

\subsection{Simulation analyses and main results}

\subsubsection{Scenario and stochastic variables} although we performed numerous simulation experiments, in this section, we only show the results obtained for a representative example. We consider a system of 30 vehicles, i.e. $m=30$, and vehicle $n=0$ evolving at an equilibrium speed $V_{\text{eq}}=V_{\max}/3$, as roughly observed in the NGSIM data set, see section~\ref{relevancess}. Vehicle car-following parameters are sampled from the joint distribution $\mathcal{T}$, as defined in section~\ref{parameters}. We introduce an acceleration perturbation to vehicle $1$, which is forced to be a non-automated vehicle. This perturbation takes the form of a PRBS input sequence of amplitude $[-1,+1]$, which remains constant over time intervals ranging from $2~\si{s}$ to $5~\si{s}$ and has a duration of $1~\si{min}$. The simulation length is set to $4~\si{min}$ as, given the considered perturbation, this is the time needed to cover all of the effects of perturbation propagation on the 30 vehicle trajectories. The stochastic variables are the sampled parameters of the 30 vehicles, the acceleration PRBS inputs, the position of the automated vehicles in the vehicle system, and the number of automated vehicles in the vehicle system. Then, we perform $25\times 4$ simulations: we repeat the simulation 25 times to consider the effects of stochastic variables; and for each of the 25 simulations we fix the seed of the introduced randomness, and consider 4 different configurations of the optimisation strategy~\eqref{LMIopt2}, e.g. different proportions of automated vehicles.  

\subsubsection{Evolution of the $\mathcal{L}_2$ norm of the speed perturbation and distribution of optimised parameters} we focus on tuning parameters $a$, $b$, and $T$ for the automated vehicles according to optimisation~\eqref{LMIopt2}, i.e. the tolerated acceleration, comfortable deceleration and safe time headway parameters. We choose $\alpha=10^3$, and work with $i=n-1$ and $j=n+2$, see Section~\ref{optimp}. 

\begin{figure}[h!]
\includegraphics[width=\textwidth]{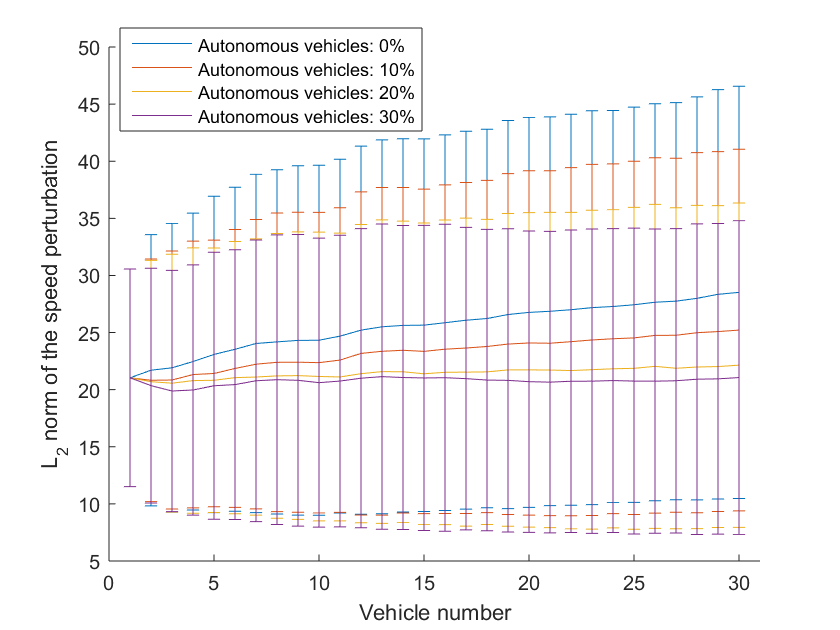}
\caption{Evolution of the $\mathcal{L}_2$ norm of the speed perturbation in the vehicle system for growing proportions of automated vehicles: $0\%$, $10\%$, $20\%$, $30\%$.}
\label{fig_error}
\end{figure} 

First, Figure~\ref{fig_error} displays the influence of the optimisation strategy~\eqref{LMIopt2} on the evolution of the $\mathcal{L}_2$ norm of the speed perturbation in the system, following the introduced PRBS acceleration inputs, for different proportion of automated vehicles. We plot the average values and errors bars of $\pm 1$ standard deviation over the $25$ simulations. The error bars show the impact of the stochastic variables over the outcome of the minimisation. The positive effects are clearly visible: an increasing percentage of automated vehicles consistently leads to lower average values and standard deviations of the $\mathcal{L}_2$ norm of the speed perturbation. Here the $(0,30)$ weak string stability condition, i.e. decrease of the $\mathcal{L}_2$ norm of the speed perturbation, is verified when $30\%$ vehicles are automated, which makes sense as the $(n-2,n+2)$ weak string stability condition of~\eqref{LMIopt2} involves a total of 4 vehicles, which means that an average of $1/4=25\%$ of automated vehicles should be enough to guarantee $(0,30)$ weak string stability provided the automated vehicles are well dispersed. Note that the PRBS input considered is actually a linear combination of step inputs of $\pm2~\si{m.s^{-2}}$, which are strong deceleration/acceleration inputs in realistic traffic. Note also that it was observed in section~\ref{lnnonln} and Figure~\ref{fig_NL3} that for such inputs the verification of the strict string stability for homogeneous traffic still leads to a decrease of the $\mathcal{L}_2$ norm of the speed perturbation.

\begin{figure}[h!]
\includegraphics[width=\textwidth]{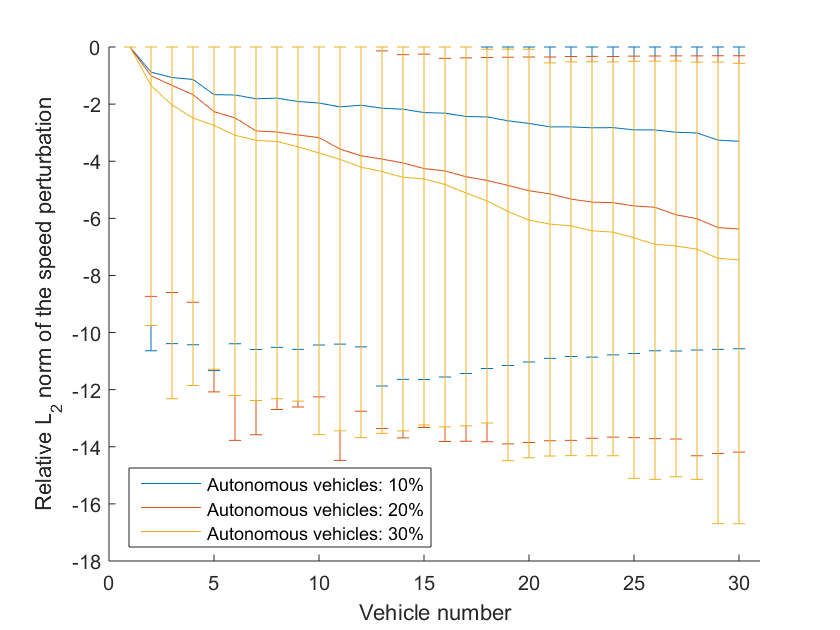}
\caption{Evolution of the relative $\mathcal{L}_2$ norm of the speed perturbation in the vehicle system for different proportions of automated vehicles: $10\%$, $20\%$, $30\%$.}
\label{fig_error2}
\end{figure} 

Second, it is worth exploring whether the proposed optimisation strategy systematically leads to positive outcomes. Figure~\ref{fig_error2} displays the average, minimum and maximum values of the deviation from the value of the $\mathcal{L}_2$ norm of the speed perturbation without any automated vehicles, for 3, 6 and 9 automated vehicles, i.e. proportions of $10\%$, $20\%$ and $30\%$. We observe that the automated vehicles with parameters derived from the optimisation strategy~\eqref{LMIopt2} contribute to systematically decrease the value of the $\mathcal{L}_2$ norm of the speed perturbation, i.e. negative relative $\mathcal{L}_2$ norms. In that sense, the proposed optimisation algorithm~\eqref{LMIopt2} consistently increases the traffic flow stability of the heterogeneous system. 

\begin{figure}[h!]
\includegraphics[width=\textwidth]{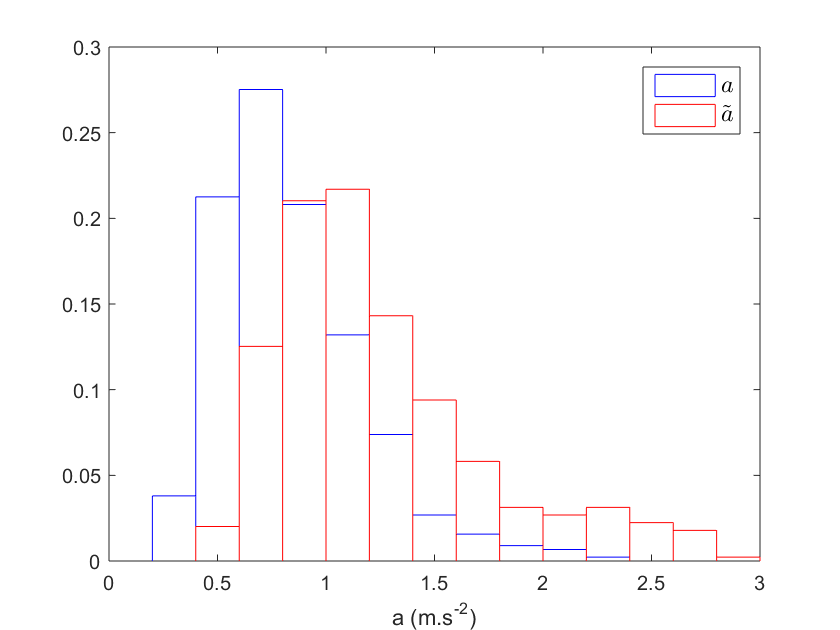}
\caption{Standard vs optimised distributions of automated vehicle parameter $a$.}
\label{fig_param1}
\end{figure} 
\begin{figure}[h!]
\includegraphics[width=\textwidth]{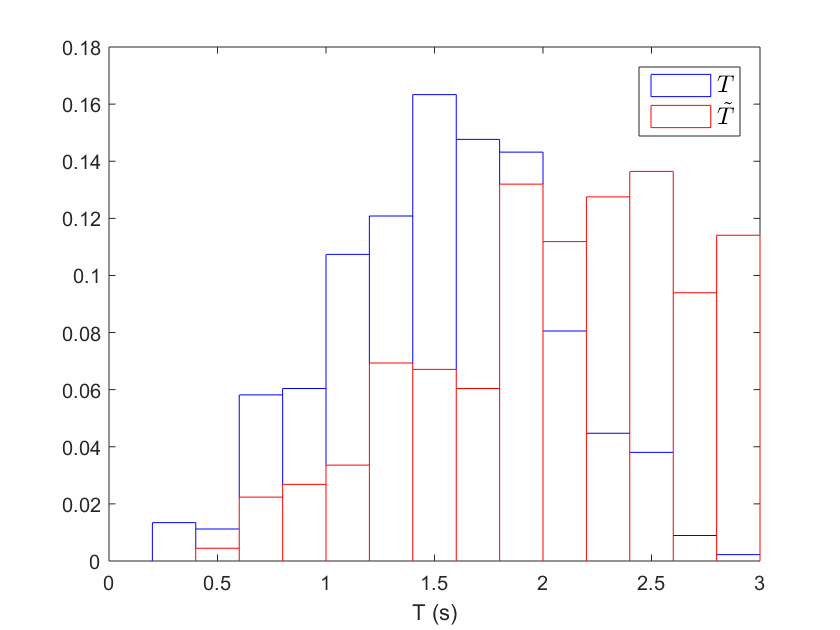}
\caption{Standard vs optimised distributions of automated vehicle parameter $T$.}
\label{fig_param2}
\end{figure} 
\begin{figure}[h!]
\includegraphics[width=\textwidth]{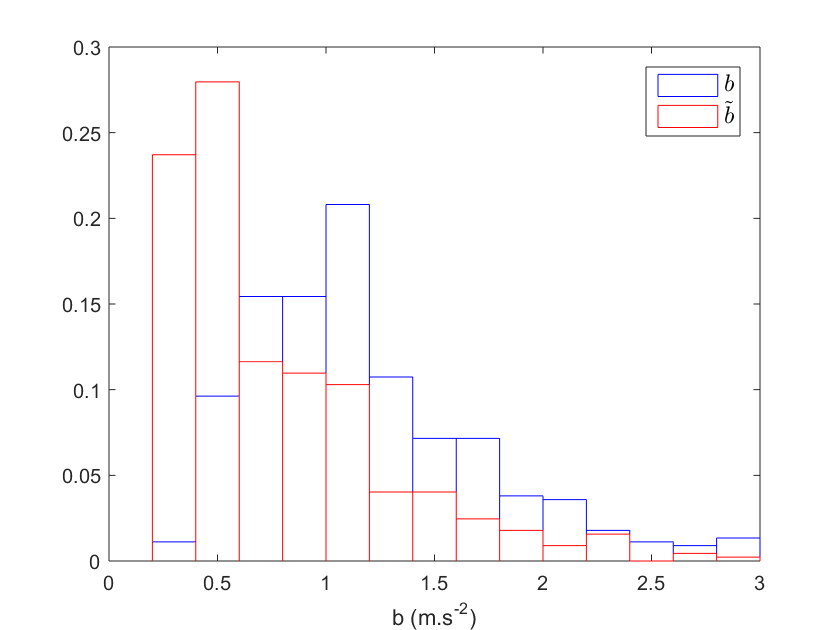}
\caption{Standard vs optimised distributions of automated vehicle parameter $b$.}
\label{fig_param3}
\end{figure} 

Finally, we look at the distributions of the optimised parameters $\tilde{a}$, $\tilde{T}$ and $\tilde{b}$, displayed in Figures~\ref{fig_param1},~\ref{fig_param2},~\ref{fig_param3}. As might be expected, the distributions are shifted but are still realistic, i.e. lead to reasonable driving behaviour by the automated vehicles. We observe that the optimisation strategy~\eqref{LMIopt2} pushes parameters $a$ and $T$ towards higher values and parameter $b$ towards lower values. This is in accordance with physical considerations: a longer safe time headway $T$ gives more time for vehicles to damp perturbations; a higher tolerated acceleration helps recover the equilibrium speed faster; a smaller comfortable deceleration results in less sharp braking and helps to smooth perturbations. 

We can also observe that parameters $b$ and $T$ are sometimes pushed towards the limits of the selected physical bounds, i.e. in this case $T=3\si{s}$ and $b=0.3\si{m.s^2}$. When this is the case, one idea to provide more flexibility to guarantee weak string stability, see Section~\ref{relaxweak}, is to increase the upper bound $T_{\text{up}}$ of the safe time headway parameter $T$, which is not a critical parameter as it does not depend on the capabilities of the vehicle or does not affect driving comfort as much as other parameters. 

\subsubsection{Systematically enforcing more stable dynamics with very few automated vehicles}

with very few automated vehicles in the vehicle system, and when the parameters of only a few leading and following vehicles are known, we might be interested in investigating how to enforce an even more stable dynamics. To do so we can introduce fictitious unstable vehicles, see section~\ref{limitation}.

\begin{figure}[h!]
\includegraphics[width=\textwidth]{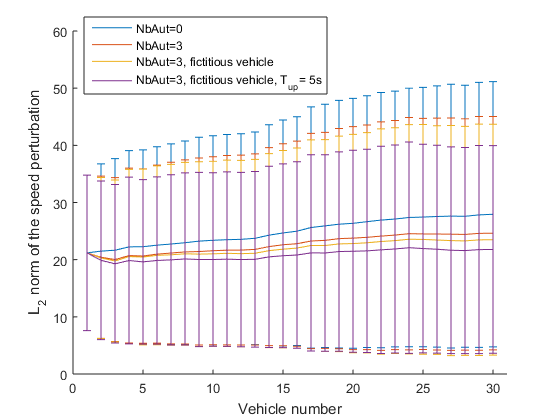}
\caption{Evolution of the $\mathcal{L}_2$ norm of the speed perturbation in the vehicle system with no automated vehicle; 3 automated vehicles and no fictitious vehicle; 3 automated vehicles, 1 fictitious vehicle and $T_{\text{up}}=3~\si{s}$; 3 automated vehicles, 1 fictitious vehicle and $T_{\text{up}}=5~\si{s}$.}
\label{fig_erroralpha}
\end{figure} 

We consider $i=n-1$ and $j=n-2$, and 3 automated vehicles in the system, i.e. $\text{NbAut}=3$. A fictitious worst case unstable vehicle $n-2$ is introduced with parameters $a_{n-2}=0.3~\si{m.s^{-2}}$, $T_{n-2}=0.3~\si{s}$, $b_{n-2}=3~\si{m.s^{-2}}$. 

Figure~\ref{fig_erroralpha} shows the results of the optimisation without and with the fictitious introduced vehicle, for 2 different admissible upper values of parameter $T_{\text{up}}$, $T_{\text{up}}=3~\si{s}$ and $T_{\text{up}}=5~\si{s}$. It is visible that adding one fictitious unstable vehicle consistently increases weak string stability in the system. However, the limitations in the stabilisation effect of automated vehicles come from the admissible set of parameter values $\Theta$. By increasing the upper bound of the admissible safe time headway $T_n$, we are able to bypass this limitation and reach $(0,30)$ weak string stability with only 3 automated vehicles in the vehicle system. This strategy may lead to less realistic and less comfortable driving behaviour, as parameters tend to move towards their upper/lower admissible bounds. However the physical bounds can be selected appropriately, as for instance parameter $T$ is less critical than parameters $a$ and $b$ in terms of vehicle capabilities and driving comfort, although increasing $T$ may encourage vehicles to change lanes and enter the empty slots created. 

Finally, the overall conclusion is that the number of automated vehicles needed to prevent perturbation growth can be reduced depending on the following parameters of the optimisation strategy: the number of vehicles for which the behavioural parameters are known, i.e. parameters $i$ and $j$, the set of admissible parameters $\Theta$, and the parameters of introduced fictitious unstable vehicles. Given a string of vehicles, a small number of automated vehicles is enough to damp the effect of realistic perturbations that would otherwise grow.

\section{Conclusion}

This paper applies $\mathcal{L}_2$ linear control theory to linearised systems of vehicles  moving according to realistic car-following models. The contributions are the following: a general framework for investigating the stability and string stability of heterogeneous traffic in the frequency domain is introduced (most previous studies assume homogeneous traffic); the definition of weak stability is introduced and its relevance in a traffic environment with a mix of automated and non-automated vehicles is highlighted; conditions for input-output stability and string stability are given for heterogeneous traffic, and for single and multiple outputs; the relation between $\mathcal{L}_2$ and $\mathcal{L}_{\infty}$ string stability is presented; the equivalence between string stability and asymptotic stability is showed not to hold for closed loop systems; simulations underline the critical feature of nonlinearities; an optimisation strategy to tune the behavioural parameters is proposed as well as its LMI formulation; the optimisation is applied to realistic data yielding promising results: a small proportion of automated vehicles, that behave similarly to their drivers, can greatly and systematically contribute to increasing traffic flow stability.

With regard to future work, (i) the impact of the non-linear dynamics, and (ii) the reasons for perturbation growth and boundedness under high acceleration inputs as the number of vehicles increases remain open questions. Regarding optimisation, (iii) the formulated LMI optimisation problem may be solved more efficiently. Finally, regarding control, (iv) the mapping of this work with online parameter identification of drivers' behavioural parameters, and the consideration of parameters uncertainty for the design of control strategies remain to be studied.

\appendix
\section{}
Singular values $\sigma_{\max}$ are defined as follows. For any $\vv{F}\in\mathbb{R}^{2\times 2}$, $$\sigma_{\max}(\vv{F}(j\omega))=\sqrt{\lambda_{\max}(\vv{F}(j\omega)^*\vv{F}(j\omega))},$$ where $\vv{F}(j\omega)^*$ is the conjugate transpose of $\vv{F}(j\omega)$ and $\lambda_{\max}(\vv{F}(j\omega)^*\vv{F}(j\omega))$ denotes the maximum of the nonzero eigenvalues of $\vv{F}(j\omega)^*\vv{F}(j\omega)$.

Following equation~\eqref{weakstr}, the product $\vv{\Gamma}_n^*(j\omega)\vv{\Gamma}_n(j\omega)$ is written: 
\begin{equation}
\vv{\Gamma}_n^*\vv{\Gamma}_n=\frac{1}{D_n^*D_n}\begin{pmatrix}
0&0\\
0&\omega^2(1+f_{n,3}^2)+f_{n,1}^2+f_{n,2}^2
\end{pmatrix},\label{appendix1}
\end{equation}
where $D_n^*D_n$ is equal to:
\begin{equation}
D_n^*D_n=\omega^4+\omega^2\left((f_{n,3}-f_{n,1})^2-2f_{n,2}\right)+f_{n,2}^2.
\end{equation}

The two eigenvalues $\lambda_1$ and $\lambda_2$ of $\vv{\Gamma}_n^*(j\omega)\vv{\Gamma}_n(j\omega)$ immediately follow:
\begin{align}
\lambda_1(\omega)&=0,\\
\lambda_2(\omega)&=\frac{\omega^2(1+f_{n,3}^2)+f_{n,2}^2+f_{n,1}^2}{\omega^4+\omega^2\left((f_{n,3}-f_{n,1})^2-2f_{n,2}\right)+f_{n,2}^2}.
\end{align}
which gives 
\begin{equation}
\sigma_{\max}(\vv{\Gamma}_n(j\omega))=\sqrt{\lambda_2(\omega)}.\label{sigmamax}
\end{equation}
We recall that, by definition, see equation~\eqref{hinf}, we have $$||\vv{\Gamma}_{n}||_{\mathcal{H}_\infty}=\sup\limits_{\omega\in\mathbb{R}}\sigma_{\max}(\vv{\Gamma}_n(j\omega)).$$ The sufficient condition for strict string stability is written $||\vv{\Gamma}_{n}||_{\mathcal{H}_\infty}\leq 1$, see equation~\eqref{submu2}, which is equivalent to $\lambda_2(\omega)\leq 1$.

Developing $\lambda_2(\omega)\leq 1$, and writing $\Omega=\omega^2$, we obtain a polynomial of order 2 in $\Omega$:
\begin{equation}
\Omega^2+\Omega(f_{n,1}^2-2f_{n,1}f_{n,3}-2f_{n,2}-1)-f_{n,1}^2\geq 0\label{poly2}.
\end{equation}
As this inequality must be verified $\forall \Omega\in\mathbb{R}_+$, we must have $f_{n,1}=0$, and the following sufficient conditions follow:
\begin{align}
f_{n,1}&=0,\\
-2f_{n,2}-1&\geq 0.
\end{align}

\section*{Acknowledgment}
This work was supported by SFI grants 11/PI/1177, 13/RC/2077 and 10/IN.1/I2980. 




%
\bibliographystyle{elsarticle-harv}
\bibliography{IEEEexample}

%




\end{document}